\documentclass{SciPost}

\hypersetup{
    colorlinks,
    linkcolor={red!50!black},
    citecolor={blue!50!black},
    urlcolor={blue!80!black}
}
\usepackage{comment}
\usepackage[normalem]{ulem}
\usepackage[bitstream-charter]{mathdesign}
\usepackage{bbm}

\DeclareSymbolFont{usualmathcal}{OMS}{cmsy}{m}{n}
\DeclareSymbolFontAlphabet{\mathcal}{usualmathcal}

\fancypagestyle{SPstyle}{
\fancyhf{}
\lhead{\colorbox{scipostblue}{\bf \color{white} ~SciPost Physics }}
\rhead{{\bf \color{scipostdeepblue} ~Submission }}

\fancyfoot[C]{\textbf{\thepage}}
}

\newcommand{\cbracket}[1]{\left\{{#1}\right\}}
\newcommand{\rbracket}[1]{\left({#1}\right)}
\newcommand{\sbracket}[1]{\left[{#1}\right]}                                    
\newcommand{\sgn}{\text{sgn}}

\newcommand{\R}{\mathbb{R}}
\newcommand{\Z}{\mathbb{Z}}

\newcommand{\N}{\mathbb{N}}
\newcommand{\dyn}{\text{dyn}}

\usepackage{physics}
\usepackage{mathtools}
\usepackage[bitstream-charter]{mathdesign}
\usepackage{tikz}
\usepackage{float}
\usepackage{cancel}
\usetikzlibrary{arrows.meta}

\begin{document}
\pagestyle{SPstyle}

\begin{center}{\Large \textbf{\color{scipostdeepblue}{
Charged moments and symmetry-resolved entanglement from Ballistic Fluctuation Theory
}}}\end{center}

\begin{center}\textbf{
     Giorgio Li\textsuperscript{1$\star$}\href{https://orcid.org/0009-0003-0196-6595}{\includegraphics[height=1em]{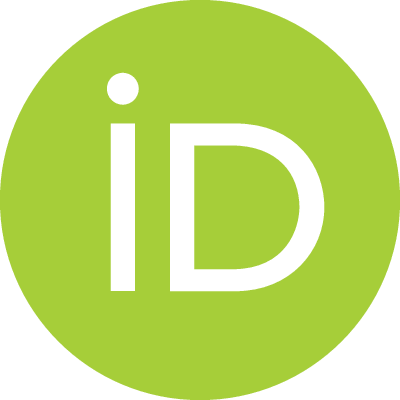}}, 
    Léonce Dupays\textsuperscript{1$\dagger$}\href{https://orcid.org/0000-0002-3450-1861}{\includegraphics[height=1em]{orcidid.eps}}, 
    Paola Ruggiero\textsuperscript{1$\ddagger$}\href{https://orcid.org/0000-0002-0891-0477}{\includegraphics[height=1em]{orcidid.eps}}
}\end{center}

\begin{center}
{\bf 1} Department of Mathematics, King’s College London, Strand WC2R 2LS, London, U.K.
\\[\baselineskip]
$\star$ \href{}{\small giorgio.1.li@kcl.ac.uk}\,,\quad
$\dagger$ \href{}{\small leonce.dupays@kcl.ac.uk}\,,\quad
$\ddagger$ \href{}{\small paola.ruggiero@kcl.ac.uk}
\end{center}

\section*{\color{scipostdeepblue}{Abstract}}
\textbf{\boldmath{%
The charged moments of a reduced density matrix provide a natural starting point for deriving symmetry-resolved R\'enyi and entanglement entropies, which quantify how entanglement is distributed among symmetry sectors in the presence of a global internal symmetry in a quantum many-body system. In this work, we study charged moments within the framework of Ballistic Fluctuation Theory (BFT). This theory describes large-scale ballistic fluctuations of conserved charges and associated currents and, by exploiting the height-field formulation of twist fields, gives access to the asymptotic behaviour of their two-point correlation functions. In Del Vecchio Del Vecchio \emph{et al.}~\cite{delVecchio_Doyon_Ruggiero_24}, this approach was applied to the special case of branch-point twist fields used to compute entanglement entropies within the replica approach. Here, we extend those results by applying BFT to \emph{composite} branch-point twist fields, obtained by inserting an additional gauge field. Focusing on free fermions, we derive analytic expressions for charged Rényi entropies both at equilibrium, in generalized Gibbs ensembles, and out of equilibrium following a quantum quench from $U(1)$ preserving pair producing integrable initial states. In the latter case, our results agree with the conjecture arising from the quasiparticle picture.
}}

\vspace{\baselineskip}


\vspace{10pt}
\noindent\rule{\textwidth}{1pt}
\tableofcontents
\noindent\rule{\textwidth}{1pt}
\vspace{10pt}
\section{Introduction}

The study of entanglement growth and saturation is central to understanding the out-of-equilibrium dynamics of many-body quantum systems \cite{Polkovnikov2011,Gogolin2016}. It provides key insights into how isolated systems approach equilibrium, with thermodynamic entropy emerging from the entanglement generated during unitary evolution \cite{Alba2019Pnas,Calabrese2018}. Entanglement dynamics also sets fundamental limits on the classical simulation of quantum systems \cite{vidal2004efficient,Verstraete2008} and serves as a probe of quantum information scrambling \cite{Sekino2008,Von_Keyserlingk2018}. 
Its importance is further emphasized by the growing ability to probe entanglement properties of many-body quantum systems experimentally, in particular with ultracold atoms\cite{Islam2015,Kaufman2016}. 

A good measure of entanglement between the subsystem $A$ and its complement $\bar{A}$ is the Von-Neumann entropy, defined as
\begin{align}
S_{EE}&=-{\rm tr}\left[\rho_{A}\ln \rho_{A}\right],\label{von_neumann}
\end{align}
where $\rho_{A}={\rm tr}_{\bar{A}}\rho$ is the reduced density matrix (RDM) and $\rho=|\Psi\rangle \langle \Psi|$ the total density matrix of the state $|\Psi\rangle$. $S_{EE}$ is often obtained as the replica limit $m\to 1$ of a larger family of entropies known as Rényi entropies (RE)
\begin{equation}
    S_m = \frac{1}{1-m}\ln \tr \rho_A^m,\label{renyi}
\end{equation}
for $m\in \mathbbm{N}$, upon proper analytic continuation to $m \in \mathbb{R}$. 

At equilibrium, one-dimensional systems allow for a number of analytical results for these quantities. For ground states of critical systems, whose low-energy physics is described by $(1+1)$ dimensional conformal field theory (CFT), partitioning the system in an interval of length $\ell$ and its complement, leads to a logarithmic behavior of the Von Neumann entropy in the subsystem size \cite{Holzhey1994,Calabrese2004,Calabrese2009}. Results are also available for gapped systems, where the famous \emph{area law} is observed \cite{Hastings2007}, as well as for integrable field theories \cite{Cardy2008}, thermal states and generalized Gibbs ensembles (GGEs) \cite{Bethe1931,Alba2017, Alba2017JSM,Mestyán2018}, and even beyond integrability \cite{Doyon2009}. 

The out-of-equilibrium dynamics of entanglement, on the other hand, is far less understood. Following a sudden quench \cite{Calabrese_Cardy_07}, the entanglement in a broad class of homogeneous one-dimensional systems is observed to grow linearly in time before saturating to its thermodynamic value \cite{Calabrese2020,Kim2013}.
While this linear growth appears to be a generic feature of nonequilibrium dynamics, its microscopic interpretation strongly depends on the underlying physical setting. In particular, only in ballistic regimes does the entanglement growth admit a description in terms of propagating quasiparticles.
This behavior can be captured using an imaginary-time path-integral approach combined with conformal invariance \cite{Calabrese2006}, and is more generally and intuitively understood in terms of the \emph{quasiparticle picture} \cite{Calabrese2005}, where entanglement is carried by ballistically propagating quasiparticles.
This picture was derived rigorously for free fermions \cite{Fagotti2008} and, more recently, recovered within Ballistic Fluctuation Theory~\cite{delVecchio_Doyon_Ruggiero_24}. Its extension to interacting integrable models requires replacing the entropy weight (accounting for both the pair-production rate and contribution to the entanglement) by the Yang–Yang entropy \cite{Yang1969}, as conjectured and supported by numerical studies in \cite{Alba2018Scipost,Alba2019Pnas}.
However, in presence of interactions, it is also known \cite{Klobas2021,Bertini2022} that the quasiparticle picture breaks down for the $m$-Rényi entropies when $m\neq 1$.
Finally, dual-unitary circuits provide an alternative realization of ballistic entanglement spreading, sharing phenomenological features with the quasiparticle picture, while not relying on integrability in the conventional sense \cite{Bertini2019,Bertini2025review}.

Both the Von Neumann \eqref{von_neumann} and Rényi entropies \eqref{renyi} can be obtained from the moments of the RDM, $Z_{m}={\rm tr}\rho^{m}_{A}$, that are computed by considering $m$ copies of the original model, ending up with a replicated theory \cite{Calabrese2004,Calabrese2009}.
Within this approach, a powerful tool are the branch point twist fields \cite{Cardy2008}, $T_{m}$ and its Hermitian conjugate $\bar{T}_{m}$. Twist fields are fields, associated to a symmetry of the Lagrangian. Branch-point twist fields are special kinds of those associated with cyclic and anticyclic permutation of the copies $j\to j+1\ {\rm mod}\ m$, under which the replicated theory is invariant by construction. Then, the moments $Z_{m}$ can be computed as correlation functions of twist fields, as pointed out in \cite{Calabrese2004,Cardy2008}. 

However, the evolution of entanglement in many-body quantum systems can be understood in a more refined way. For systems that possess an internal global symmetry, such as a $U(1)$ symmetry, it is interesting to look at how the entanglement decomposes in different symmetry sectors. This led to the notion of \emph{symmetry-resolved} entanglement entropy \cite{Laflorencie2014,Goldstein_Sela_18,Xavier2018,Lukin2019}, which has attracted considerable interest in recent years
(see Ref.~\cite{Castro_Alvaredo_2024} and references therein for a review of the main ideas and techniques, as well as a comprehensive survey of the literature). 
A useful starting point to compute the latter are the so-called \emph{charged moments} $Z_{m}(\alpha)={\rm tr} \left(\rho^{m}_{A}e^{i\alpha Q_{A}} \right)$ where $Q_{A}$ is the $U(1)$ charge restricted to the subsystem $A$. As first noticed in \cite{Goldstein_Sela_18}, they can be understood as a generalization of $Z_m$ corresponding to the insertion of an Aharonov–Bohm flux $\alpha \in \mathbb{R}$ conjugate to $Q_A$. The symmetry-resolved entanglement brings supplementary information on the many-body system. From it, one can infer the entanglement spectrum of each charge block of the reduced density matrix \cite{Calabrese2008,Goldstein_Sela_18}. It also allows to define the so-called \emph{entanglement asymmetry}, which measures how much a local symmetry is broken and is then restored in dynamical settings \cite{Ares2023}. Charged moments have also been studied in the holographic setting \cite{Belin2013,Caputa2016} and in the study of entanglement of mixed states \cite{Cornfeld2018,Shapourian2019}. 

In this article, we study the charged moments $Z_m(\alpha)$, both at equilibrium—in generalized Gibbs ensembles (GGEs)—and out of equilibrium following a quantum quench, by employing the recently developed theory of Ballistic Fluctuation Theory (BFT) \cite{Myers_Doyon_19, Myers_Bhaseen_Harris_Doyon_20}.
BFT can be viewed as a generalization of the relation between the partition function and the free energy in thermodynamics. More specifically, it provides a hydrodynamic large-deviation description of fluctuations of conserved charges through their associated two-currents (densities and currents), integrated along arbitrary rays in space–time, in homogeneous and stationary states. In this work, we focus on horizontal and vertical rays, which suffice for our purposes and encode the full counting statistics (FCS) of conserved charges in a spatial subsystem and of time-integrated currents, respectively. Within BFT, these quantities arise as different limits of a unified fluctuation framework.
Owing to the height-field formulation of twist fields, which links them to the FCS of charges and currents, BFT also provides access to two-point correlation functions of twist fields, as shown in Ref.~\cite{delVecchio_Doyon_Ruggiero_24} for Rényi entropies. Here, we extend those results to charged moments, applying BFT to the calculation of correlation functions of \emph{composite} branch-point twist fields, obtained by including an additional flux $\alpha$, as mentioned above \cite{Goldstein_Sela_18}.
In this work, we focus on free fermionic models, where the BFT framework allows for explicit and fully controlled calculations.
Our results agree with existing literature on the time evolution of Rényi entropies based on exact free-fermion calculations \cite{Parez2021, Parez_Bonsignori_Calabrese_Long}, and provide a rigorous derivation of charged moments for a broader class of initial states.

The paper is organized as follows. Sec.~\ref{sec:technical_introduction} introduces charged moments and the system under consideration: free fermions. Sec.~\ref{sec:symmetries_and_twist_fields} introduces branch-point twist fields as the local generators of a symmetry and their link to currents. Sec.~\ref{equilibrium_case} discusses the results for the charged moments at equilibrium. Sec.~\ref{out_of_equilibrium} introduces the notion of integrable quench and, in particular, the class of initial states of interest when studying symmetry resolution of entanglement, explaining how BFT is used to treat out-of-equilibrium scenarios. In Sec. \ref{sec:symm_res_Fourier_transform} we briefly comment on the symmetry resolved entropies, obtained via Fourier transform. Sec.~\ref{conclusion} concludes our findings. Technical details of the calculations can be found in the appendices. 

\section{Technical introduction: definitions \label{sec:technical_introduction}}

In this section, we provide additional context and precise definitions of the concepts introduced above and the systems of interest. 

\subsection{Definition of the charged moments and symmetry-resolved entropies}
Consider a system that possesses a symmetry, associated to a conserved charge $Q$.
Though the
physical picture is expected to hold for a generic local charge, with some technical complications, we assume $Q$ to be ultra-local\footnote{A charge $Q = \int \dd x q(x)$ is ultra-local if its density $q(x)$ is supported at a single point $x$.} for simplicity. This property implies an exact decomposition of the form
$Q=Q_A\otimes \mathbbm{1}_{\bar{A}}+\mathbbm{1}_{A}\otimes Q_{\bar{A}}$, for any arbitrary subset $A$ where $Q_A$ is $Q$ restricted to the subsystem $A$, and $\mathbbm{1}_A$ the identity operator in that subset.
If $\ket{\psi}$ is an eigenstate of $Q$, the total density matrix $\rho$ commutes with the charge, $\sbracket{\rho,Q}=0$. This also implies $\sbracket{\rho_A,Q_A}=0$. As a result, $\rho_A$ has a block-diagonal structure, meaning that it can be decomposed into contributions from different charge sectors \cite{Goldstein_Sela_18}
\begin{equation}\label{eq:RDM_charge_sector}
    \rho_A  = \bigoplus_q \Pi_q \rho_A= \bigoplus_q p_q \rho_A(q),
\end{equation}
where $\Pi_q$ is the projector on the eigenspace of $Q_A$ with eigenvalue $q$. The term $p_q = \tr (\Pi_q\rho_A)$ represents the probability of obtaining the value $q$ when a measurement of the restricted charge $Q_A$ is performed. For instance, for the number operator $N_A$, which counts the number of particles in the subsystem $A$, the projector is a delta function $\delta({N}_{A}- n_{A})$, with $n_A$ being the eigenvalue. 

Based on this charge-sector decomposition, we define the symmetry-resolved Rényi entropy (SRRE) of order $m \in \mathbb{N}$, for each eigenvalue, as $q$
\begin{equation}
    S_m(q)=\frac{1}{1-m}\ln\tr\rho_A^m(q).
\end{equation}
The SREE is then obtained as the limit of the Rényi-entropies $S(q)=\lim_{m\to1}S_m(q)$. 

The total entanglement entropy can be divided in two distinct parts \cite{Wiseman2003,Lukin2019, Xavier2018}
\begin{equation}
    S_{EE}=\sum_q p_q S(q)-\sum_q p_q \ln p_q \equiv  S^c+S^f. \label{symmetry_resolved_entanglement}
\end{equation}
$S^c$ denotes the configurational entanglement entropy, namely the probability-weighted average of the entanglement entropy within each charge sector.
$S^f$ denotes the fluctuation entanglement entropy, which measures the entropy due to the fluctuations of the charge's value within subsystem $A$. 
For instance, when the conserved charge is the particle-number operator, the configurational entanglement reflects how particles are arranged and correlated within the subsystem A. By contrast, the fluctuation entanglement entropy is governed by particle-number fluctuations across the bipartition and is therefore sensitive to the transport properties of the system, and tend to saturate after reaching the equilibrium or if transport is suppressed by localization~\cite{Lukin2019}.

To facilitate the calculation of SRRE, it is customary to introduce the moment generating function or \emph{charged moments}
\begin{equation} \label{def:chsrgedmoments}
    Z_m(\alpha)=\tr \left( \rho_A^m e^{i\alpha Q_A} \right) \ .
\end{equation}
They generate the moments of $Q_A$ through
\begin{equation}
    \frac{1}{i^n}\pdv[n]{\alpha}Z_m(\alpha)=\tr \left( \rho_A^m Q_A^n \right).
\end{equation}
The Fourier transform of the charged moments gives an unormalized probability distribution
\begin{equation} \label{moments_fourier}
    \mathcal{Z}_m(q)=\int_{-\pi}^{\pi} \frac{\dd \alpha}{2\pi}Z_m(\alpha)e^{-i\alpha q}.
\end{equation}
In particular, for $m=1$, one obtains the probability distribution $\mathcal{Z}_1(q)=p_q$, which is the Fourier transform of the characteristic function.

The SREE is related to the Fourier transform of the charged moments through the relation~\cite{Goldstein_Sela_18}
\begin{equation}
    S_m(q)=\frac{1}{1-m}\ln\left[\frac{\mathcal{Z}_m(q)}{\mathcal{Z}^{m}_{1}(q)}\right],\label{symmetry_resolved}
\end{equation}
including normalization.
\subsection{Systems of interest: free fermions}\label{sec:model}
The focus of this article is a one-dimensional free fermionionic system, governed by a local, translationally invariant Hamiltonian. 
Throughout, we work in the hydrodynamic coarse-grained limit~\cite{spohn2012large}, in which microscopic degrees of freedom are replaced by locally averaged fields defined over mesoscopic cells.
Consequently, without loss of generality, we describe lattice and continuum systems in terms of a single coarse-grained fermionic field $\psi(x)$. The field $\psi(x)$, defined on real space $x\in\R$, satisfies the canonical anticommutation relations
\begin{equation}
    \{\psi(x),\psi^\dagger(y)\}\equiv\psi(x)\psi^\dagger(y)+\psi^\dagger(y)\psi(x)=\delta(x-y),\\
\end{equation}
and
\begin{equation}
    \{\psi(x),\psi(y)\}=0,\quad 
    \{\psi^\dagger(x),\psi^\dagger(y)\}=0.
\end{equation}
We denote by $\psi(k)$ the Fourier modes associated with $\psi(x)$; the two representations are related by the Fourier transform
\begin{equation}\label{eq:field_real_space}
    \psi(x)=\int  \frac{d k}{2 \pi}\psi(k)e^{ikx},\quad  \psi(k)=\int_{}^{} \dd x e^{-ikx}\psi(x) ,
\end{equation}
where $k$ is the momentum. In momentum space the Hamiltonian is diagonal and takes the form 
\begin{equation}
    H = \int_{}^{} \frac{\dd k}{2 \pi} E_k \psi^\dagger(k)\psi(k) ,
\end{equation}
where the single particle dispersion $E_k$ is assumed to be an even function of $k$.
\section{Charged moments and twist fields \label{sec:symmetries_and_twist_fields}}
The computation of the charged moments $Z_m(\alpha)$ for $m\in\mathbb{N}$
can be mapped to the evaluation of a correlation function of $\emph{twist fields}$. Let the subsystem $A$ be a single interval such as $A=\sbracket{x_1,x_2}$, then $Z_m (0) = \tr \rho_A^m$ has the path-integral interpretation of a partition function on a Riemann surface geometry with a cut along the interval $\sbracket{x_1,x_2}$, connecting the different Riemann sheets. The boundary conditions across the cut can be implemented by the insertion of two twist operators at the end-points of the interval $A$ in the \emph{replica model} \cite{Calabrese2004, Cardy2008}. This gives 
\begin{equation}
    Z_m (0)= \expval{T_m(x_1)\bar{T}_m(x_2)}_{\rho^{\otimes m}},
\end{equation}
where $T_m(x)$ is known as the \emph{branch-point} twist field, $\bar{T}_{m}(x)$ denotes its inverse, and the expectation value is on $\rho^{\otimes m} = \otimes_{j=1}^m \rho_j$, with $\rho_j$ the original density matrix in copy $j$. This twist field operator is associated with the cyclic permutation symmetry of the initial partition function $Z_m(0)$. 

Similarly, for $\alpha \neq 0$, $Z_m (\alpha)$ can be written as a two-point correlation function of a \emph{composite} branch-point twist field $T_{m,\alpha}(x)$~\cite{Goldstein_Sela_18}
\begin{equation} \label{Zalpha_twist}
    Z_m (\alpha) =\expval{T_{m,\alpha}(x_1)\bar{T}_{m,\alpha}(x_2)}_{\rho^{\otimes m}} \ .
\end{equation}
Here the path-integral interpretation is that of a partition function on the same Riemann surface modified with a space-time Aharonov-Bohm flux $\alpha$ insertion~\cite{Goldstein_Sela_18}. 

To compute this two-point correlation function, we exploit the relation between twist fields and conserved charges, which follows from their role as generators of symmetry transformations.

\subsection{Height field formulation of twist fields}
The twist fields introduced above admit an explicit representation called height-field formulation. This comes from their property of semi-local generators of a symmetry. Consider a generic charge $Q$ generating a symmetry of the model.
The symmetry action on the field $\psi (x)$ is then given by
\begin{equation} \label{eq:symmetry_transformation}
    \check{\psi}(x)=e^{i\eta Q}\psi(x)e^{-i\eta Q}, \quad \forall \eta,x \in \R.
\end{equation}
Assuming that $Q$ is extensive and can be written as an integral over a local density $q(x)$, one can define the associated \emph{height field} $\phi(x)$ as 
\begin{equation}\label{eq:height_field}
    \phi(x)=\int_{x}^{\infty} \dd x' q(x'),
\end{equation}
and the exponential of such field
\begin{equation}\label{eq:height_field_twist_field}
    T_Q(\eta; x)=\exp \left[i\eta \phi(x)\right],
\end{equation}
can be shown to implement the following equal-time exchange relations
\begin{equation}\label{eq:equal_time_exchange_relations}
    T_Q(\eta, x)\psi(y)=\begin{cases}
        \check{\psi}(y) T_Q(\eta; x) & y\geq x,\\
        \psi(y) T_Q(\eta; x) &  y<x \ .
    \end{cases}
\end{equation}
Similar exchange relations follow for its inverse $T_Q(\eta; x)^{-1} = T_Q(\eta; x)^{\dagger} = T_Q(- \eta; x)=\bar{T}_Q(\eta; x)$.
At this stage, the time dependence is omitted as only the equilibrium case is of interest.
Since Eq.~\eqref{eq:equal_time_exchange_relations} are the same exchange relations associated with twist fields~\cite{Calabrese2004,Cardy2008}, Eq.~\eqref{eq:height_field_twist_field} is known as the \emph{height field formulation} of the twist field $T_Q (\eta; x)$. 
Expectation values of $T_Q (\eta ; x)$ generate the Full Counting Statistics of the charge $[x, \infty )$, providing a direct connection to the regime where BFT applies.

One can define a twist field via Eq.~\eqref{eq:height_field_twist_field} for any local and extensive charge $Q$~\footnote{Locality and extensivity of $Q$ ensures $T_Q(\eta, x)$ to be a local operator and to have a non-trivial cumulative effect on the system, respectively.}.
If then $Q$ acts diagonally (i.e., generates a U(1) symmetry) in the eigenbasis of the field $\psi(x)$, the action \eqref{eq:symmetry_transformation} becomes
\begin{equation} \label{eq:phase-action}
\check{\psi}(x) = e^{- i \eta h} \, \psi(x),
\end{equation}
where $h$ is the charge of the field (one-particle eigenvalue of $Q$).
In this case, the exchange relations \eqref{eq:equal_time_exchange_relations} simplifies to
\begin{equation}\label{exchange_relation_U1}
    T_Q(\eta; x)\psi(y)=
    \begin{cases}
        e^{-i \eta h }\psi(y) T_Q(\eta;  x) & y\geq x,\\
        \psi(y) T_Q(\eta; x) & y<x \ .
    \end{cases}
\end{equation}

This is the case for the number operator $N$. In fact, the corresponding twist field, that we denote as
\begin{equation} \label{eq:Twist_varphi_Nx}
    T(\eta; x) = e^{i \eta \varphi (x)}, \qquad  \quad    \varphi (x) = \int_x^{\infty} \dd x' n(x'),
\end{equation}
with $n(x) = \psi^\dagger(x)\psi(x)$ being the real-space density,
reproduces \eqref{exchange_relation_U1} with one-particle eigenvalue $h=1$.
\subsection{Branch-Point Twist Fields and Vertex Fields in the replica theory \label{sec:branch_point_vertex} }
This subsection establishes the connection between branch-point twist fields and the height field formulation, which was first shown in~\cite{delVecchio_Doyon_Ruggiero_24}. Then extends it to include the insertion of a flux.  

In the replica method used to compute $Z_m (0)= \tr \rho_A^m$, one considers $m$ copies of the original theory \cite{Calabrese2004}. This multi-copy theory can be constructed by considering $m$ copies of the field $\psi(k)$, which is done by adding a label to the field as $\psi(k)\to \psi_j(k)$, where $j\in \cbracket{1,\ldots,m}$ denotes the copy. These fields are now assumed to satisfy the fermionic anticommutation relations~\footnote{Note that the mapping of the original quantities (cf.~\eqref{def:chsrgedmoments}) to the replicated theory does not uniquely determine the commutation relations between different copies, so that one is free to adopt, for instance, commuting or anticommuting ones. However, once this choice is made, it must be applied consistently throughout any manipulations, such as the evaluation of multi-copy operator products.
We consider anticommuting replicas here because we will need to change basis to diagonalize the twist field action, and, by starting with anticommuting fermions, the new fields will remain with well-defined fermionic statistics.}
\begin{align}
    &\{\psi_i(k),\psi^\dagger_j(k')\}=2\pi\delta_{ij}\delta(k-k'),\\
    &\{\psi_i(k),\psi_j(k')\}=\{\psi^\dagger_i(k),\psi^\dagger_j(k')\}=0\quad \forall i,j \in \cbracket{1,\ldots,m}.
\end{align}
By construction any replica theory possesses a cyclic-permutation symmetry, $\Z_m$. The \emph{branch-point} twist fields are the operators associated with this symmetry and they satisfy the following equal-time exchange relations, 
\begin{equation}
    T_m(x)\psi_j(y)=\begin{cases}
        \psi_{j+1}(y)T_m(x),\quad y\geq x,\\
        \psi_{j}(y)T_m(x),\quad y< x,
    \end{cases}
\end{equation}
with the following boundary conditions $\psi_{m+1}(x)=-\psi_1(x)$~\cite{Cardy2008}. However, due to the free nature of the fields, the multi-copy theory acquires an enhanced $U(m)$ symmetry (this is not the case for generic interacting integrable systems). It is then possible to embed this $\Z_m$ symmetry in the larger symmetry.
More precisely, the $\Z_m$ group of interest can be seen as part of a  $\sbracket{U(1)}^m \subset U(m)$ symmetry subgroup associated to the basis that diagonalizes $T_m(x)$, as we now explain. 

The diagonalization of the branch-point twist fields action is obtained through the Fourier-transform $\mathcal{F}_{j \to p}$ of the copy (replica) index. 
\begin{equation}\label{eq:fourier_transform_copy}
    \tilde{\psi}_p(x)= \mathcal{F}_{j \to p} [\psi_j (x)] \equiv \frac{1}{\sqrt{m}}\sum_{j=1}^{m}e^{i\lambda_pj}\psi_j(x),
\end{equation}
where 
\begin{align}
    \lambda_p &= \frac{\pi(2p-1)}{m}, \quad p\in F_{m}=\cbracket{-\frac{m}{2}+1,-\frac{m}{2}+2,\ldots,\frac{m}{2}}. \label{eq:fermionic_eigenvalues}
\end{align}
There is a number operator and density associated with each Fourier copy $p$
\begin{equation} \label{number_p}
    \tilde{N}_{p} = \int \dd x' \tilde{n}_p (x') , \qquad  \tilde{n}_p (x) = \tilde{\psi}^\dagger_p(x)\tilde{\psi}_p(x) =  \mathcal{F}_{j \to p}[n_j (x)=\psi^\dagger_j(x)\psi_j(x) ].
\end{equation}
The branch point twist fields $T_m(x)$ acts diagonally on $\tilde{\psi}_p(x)$
\begin{equation}\label{eq:twist_field_on_fourier_basis}
    T_m(x)\tilde{\psi}_p(y)=
    \begin{cases}
        e^{-i\lambda_p}\tilde{\psi}_{p}(y)T_m(x),\quad y\geq x,\\
            \tilde{\psi}_{p}(y)T_m(x),\quad y< x,
    \end{cases}
\end{equation}
See Appendix~\ref{sec:diagonalization_branch_point_twist_field} for a detailed derivation. 

We now would like to determine the height-field formulation of the twist field $T_m(x)$. To this aim, we note that in this basis, there is a $\sbracket{U(1)}^m$ symmetry subgroup associated to the transformation $\tilde{\psi}_p \to  e^{i \eta_p } \tilde{\psi}_p$, with $\eta_p \in \mathbb{R}$, for each Fourier-copy $p$. The twist field associated to such symmetry is
\begin{equation} \label{T_vec_lambda_x}
    \tilde{T}(\vec{\eta};x) = \exp \left(i  \sum_{p\in F_m} \eta_p \tilde{\varphi}_p(x)\right),\quad
    \vec{\eta}=(\eta_p)_{p\in F_m}, 
\end{equation}
written in the height field formulation, with height field 
\begin{equation}
    \tilde{\varphi}_p(x) = \int_{x}^{\infty} \dd x' {\tilde{n}}_p (x'),
\end{equation}
with $\tilde{n}_p (x) $ defined in Eq.~\eqref{number_p}. 
$\tilde{T}(\vec{\eta};x)$ verifies the exchange relation~\eqref{exchange_relation_U1} for $U(1)$ charges, with charge eigenvalue $h=1$, and $\eta_{p}$ the height-field parameter. Comparing Eq.~\eqref{eq:twist_field_on_fourier_basis} and Eq.~\eqref{exchange_relation_U1}, it becomes apparent that the height field formulation of $T_{m}(x)$ is given by Eq.~\eqref{T_vec_lambda_x} for the specific values $ \eta_p =\lambda_p$ (cf. \eqref{eq:fermionic_eigenvalues}). Hence, we define $\vec{\lambda}_m=(\lambda_p)_{p\in F_m}$ the vector such that 
\begin{equation} \label{eq:T_lambda_n}
\tilde{T}(\vec{\lambda}_m;x)=T_m(x).   
\end{equation}
This embedding is important as it allows to consider twist fields associated to a \emph{continuous} symmetry $[U(1)]^m$ while the initial replica symmetry $\Z_m$ is \emph{discrete}. And this fact in turn will be essential for the BFT flow to be well defined \cite{Myers_Doyon_19}.

For the purpose of computing the charged moments in Eq.~\eqref{def:chsrgedmoments}, we would like to extend this construction to include the insertion of a flux \cite{Goldstein_Sela_18}.
The mapping between original operator defined in the single-copy theory and the one in the replica theory is given by
\begin{equation}
\exp(i\alpha Q_A) \;\;\longrightarrow\;\; 
\exp\left(i \frac{\alpha}{m} \textstyle \sum_{j=1}^m Q_{A,j}\right),
\end{equation}
where we introduced the restricted charge operator $Q_{A,j}$ for each copy $j \in \{1, \cdots,  m \}$.
The total flux $\alpha$ in the initial theory can be interpreted as the insertion of a flux $\alpha/m$ for each replica $j$ (this choice also preserves the $\Z_n$ symmetry).

Focusing on the particle number, $Q = N$, we start by defining the \emph{vertex field} as 
\begin{equation}\label{eq:V_lambda_alpha}
    V_\alpha(x) = \exp \left(i \frac{\alpha}{m} \sum_{j=1}^m \varphi_j(x)\right), 
    \qquad \varphi_j (x) = \int_x^{\infty} \dd x' n_j (x') \ , 
\end{equation}
and we denote its inverse as $V_\alpha^{-1}=\bar{V}_\alpha$. 
It is then immediate that for, e.g., $A=\sbracket{x_1, x_2}$
\begin{equation} \label{FCS_V_Vbar}
V_{\alpha} (x_1) \bar{V}_{\alpha} (x_2) = \exp\left(\frac{i\alpha}{m} \textstyle \sum_{j=1}^m N_{A,j}\right) ,
\end{equation}
and this reproduces exactly the insertion of $\exp(i \alpha N_A)$ when traced over replicas.

Additionally, from \eqref{eq:V_lambda_alpha}, it is evident that $V_\alpha(x)$ is a twist field with diagonal action in the eigenbasis of $\psi_j (x)$, i.e.
\begin{equation}\label{eq:equal_time_exchange_vertex_field}
    V_\alpha(x)\psi_j(y) = \begin{cases}
        e^{-i\alpha/m}\psi_j(y)V_\alpha(x),\quad y\geq x,\\
        \psi_j(y)V_\alpha(x),\quad y<x.
    \end{cases}
\end{equation}
But more importantly, its action remains diagonal in the Fourier basis $\tilde{\psi}_p$, where its form remains unchanged, s.t.
\begin{equation}
    V_\alpha(x)=\tilde{T}\left(\vec{\frac{\alpha}{m}}; x \right), \quad \vec{\frac{\alpha}{m}} =
    \frac{1}{m}(\alpha,\ldots,\alpha).
\end{equation}
This is because the coupling fields in $\vec{\frac{\alpha}{m}}$ are $p$-independent. 

Finally one can construct the composite twist field $T_{m,\alpha}(x)$ implementing both the $[U(1)]^{m}$ symmetry and the flux insertion in a diagonal way as
\begin{equation} \label{composite_twist_field}
    T_{m,\alpha}(x) =  \tilde{T}\rbracket{\vec{\lambda}_{m,\alpha};x}
\end{equation}
with
\begin{equation} \label{moments_FCSs}
    \vec{\lambda}_{m,\alpha} \equiv \vec{\lambda}_m+\frac{\vec{\alpha}}{m}.
\end{equation}
Putting everything together, we have explicitly constructed the twist fields $T_{m,\alpha}$ in \eqref{Zalpha_twist} using the height field formulation. Hence, the 2-point 
correlation function is reformulated in terms of the FCS of $U(1)$-charges. This reads
\begin{equation} \label{mapping_to_FCS}
     Z_m (\alpha)
    = \expval{\exp \left(i \textstyle \sum_{p\in F_m} \lambda_{p,\alpha} \, \tilde{N}_{A,p}\right)}_{\rho^{\otimes m}}, \qquad \lambda_{p, \alpha}= \lambda_p + \frac{\alpha}{m} 
\end{equation}
and  
$\tilde{N}_{A,p}$ is the number operator~\eqref{number_p} restricted to subsystem $A$.
\section{Charged Moments at Equilibrium \label{equilibrium_case}}
In the previous section, we demonstrated that the computation of charged moments can be mapped onto a full counting statistics problem for the particle-number operator restricted to a subsystem 
$A$. While FCS problems are generally difficult to solve, considerable simplifications arise in equilibrium, in particular, when the subsystem becomes asymptotically large, $A\to\infty$, and the large-deviation principle applies~\cite{Gartner_77,Ellis_84}. In this regime, the fluctuations are governed by standard thermodynamics, as can be understood within the framework of BFT~\cite{Myers_Doyon_19}~\footnote{Note that BFT concerns the FCS of charge densities and currents (two-currents), we specialize the result to horizontal space-time paths.}.

Below, using the mapping in Eq.~\eqref{mapping_to_FCS}, we compute the charged moments at equilibrium in a GGE. In Sec.~\ref{GGE}, we introduce the concept of GGE, which naturally arises in one-dimensional systems possessing an infinite set of conserved quantities. In Sec.~\ref{sec:large_deviation_theory_flow_equation}, we explain how the FCS of a conserved quantity $Q$ in a large subsystem $A$ can be evaluated, using BFT. Note that, while the approach is more general, we will only give explicit closed-form expressions valid for free fermionic systems only.
Then, in Sec.~\ref{sec:scgf_at_equilibrium}, we relate the asymptotic form of the FCS to the charged moments.

\subsection{Generalised Gibbs ensembles \label{GGE}}
GGEs are the maximum-entropy states (MES) to which an isolated quantum system relax at asymptotically long times. It generalizes the well-known Gibbs ensemble~\cite{Jaynes_57} by incorporating additional conserved quantities beyond the Hamiltonian. 
In particular, one-dimensional free systems possess an infinite set of conserved charges, which we denote by $Q_\mu$ with $\mu \in \mathbb{N}$. These charges commute with the Hamiltonian,
\begin{equation}
[H,Q_\mu] = i\partial_t Q_\mu = 0.
\end{equation}
For free fermionic systems, this constraint fixes the charges to be diagonal in the particle-number basis. They can therefore be written as
\begin{equation}
    Q_\mu = \int_{}^{}\frac{\dd k}{2\pi} h_\mu(k)N_k,\qquad N_k = \psi^\dagger(k) \psi(k),\label{decomposition_of_the_charge}
\end{equation}
where $h_\mu(k)$ is the single-particle eigenvalue associated with the conserved charge $Q_\mu$. 
Given this infinite set of conserved quantities, one can construct thermal states and, more generally, generalized Gibbs ensemble (GGE) which are characterized by a \emph{generalized Boltzmann weight} $w_k = \sum_\mu \beta_\mu h_\mu(k)$, here the $\beta_\mu$ play the role of Lagrange multiplier. The state can be written as~\cite{Kormos_Shashi_Chou_Caux_Imambekov_13,Palmai_Konik_18}
\begin{equation}
    \rho_{w} = \frac{e^{-\int_{}^{}  \frac{\dd k}{2\pi}w_k N_k}}{\tr\rbracket{e^{-\int_{}^{}  \frac{\dd k}{2\pi}w_k N_k}}}
\end{equation}
We denote expectation values with respect to $\rho_w$ by
\begin{equation}\label{expval_w}
    \expval{\ldots}_w=\tr\rbracket{\ldots \rho_w}.    
\end{equation}
For each conserved charge $Q_{\mu}$ there is a density $q_\mu(x)$ associated with it such that $Q_\mu = \int_{}^{} q_\mu(x) \dd x$. 
Its expectation value in a GGE is given by
\begin{equation}
    \expval{q_\mu (x)}_w=
    \int_{}^{}\frac{\dd k}{2\pi} h_\mu(k) n_k,
\end{equation}
The result is independent of $x$ as $\rho_w$ is a translationally invariant state. The function $n_k$ is the thermodynamic occupation function defined as 
\begin{equation}
    n_k=\expval{N_k}_{{\color{red}}w}=\frac{1}{1+ e^{w_k}}.\label{eq:n_k_GGE}
\end{equation}
\subsection{FCS and SCGF 
of conserved charges
}\label{sec:large_deviation_theory_flow_equation}

Let us consider a generic interval $A$ and the conserved charge $Q_A$ restricted to it. We are interested in the expectation value $\expval{\exp (\eta Q_A)}_w$ in a GGE (cf.\eqref{expval_w}), where, for the moment, we assume $\eta \in \mathbb{R}$.
Owing to the homogeneity of the GGE, the particular position of the interval is not important and this quantity only depends on the length of the interval, which we denote in this section by $|A|$. The fluctuations of $Q_A$ in a given state are fully characterized by the scaled cumulants
\begin{equation}
    c_j = -\lim_{\abs{A}\to\infty}\frac{1}{\abs{A}}\expval{[Q_A]^j}_w^c,\quad j\in\N,
\end{equation}
where $\expval{\ldots}^c$ denotes connected correlation functions. The scaled cumulants are generated by the so-called Scaled Cumulant Generating Function (SCGF)
\begin{equation}
    G(\eta) = \sum_{j=1}^\infty \frac{\eta^j}{j!} c_j.
\end{equation}
Formally, such function is defined as
\begin{equation} \label{def:limit_SCGF}
    G(\eta)=-\lim_{\abs{A}\to\infty} \frac{1}{\abs{A}}\ln \expval{e^{\eta Q_A}}_w,
\end{equation}
which implies the following asymptotic form of the FCS
\begin{equation}\label{eq:scaled_cumulant_generating_function}
    \expval{e^{\eta Q_A}}_w\asymp e^{-\abs{A}G(\eta)},
\end{equation}
where the notation $f(x)\asymp g(x)$ means that $\lim_{x\to\infty}\ln [f(x)]/\ln [g(x)]=1$. 
If $G(\eta)$ is finite and differentiable, then the Gärtner–Ellis theorem ensures that the distribution of $Q_A$ satisfies a large deviation principle, with linear scaling~\cite{Gartner_77,Ellis_84}.

Then, the probability of obtaining $q_A$ when a measurement of $Q_A$ is performed (cf.~\eqref{eq:RDM_charge_sector}) takes the form
\begin{equation}\label{eq:large_deviation_principle}
   p(q_A) \asymp e^{-\abs{A}I(q_A)},
\end{equation}
where $I(q_A)$ is called \emph{rate function}, related to $G(\eta)$ by the Legendre-Fenchel transform
\begin{equation}
   I(q_A) = \sup_{\eta\in\R}\cbracket{\eta q_A-G(\eta)}.
\end{equation}
Let $\rho_{\varepsilon_\eta}$ be a GGE $\rho_w$ biased by $e^{\eta Q}$
\begin{equation}\label{eq:equilibrium_biased_gge}
    \rho_{\varepsilon_\eta}=\frac{e^{\eta Q}\rho_w}{\tr \left( e^{\eta Q}\rho_w \right)},
\end{equation}
where $\eta$ plays the role of an additional Lagrange multiplier. 
The generalized Boltzmann weight associated with this biased ensemble in the case of free fermionic systems, 
is given by
\begin{equation}
    \varepsilon_\eta = w_k-\eta h(k),
\end{equation}
where $h(k)$ is the single particle eigenvalue associated to $Q$. It was shown in~\cite{Myers_Doyon_19} (see also references therein) that $G(\eta)$ could be written as the difference of free energy densities
\begin{equation}
    G(\eta)=f\sbracket{w-\eta h}-f\sbracket{w}, \label{SCGF}
\end{equation}
where the fermionic free energy density function takes the form
\begin{equation}
    f\sbracket{w}=- \int_{}^{} \frac{\dd k}{2\pi} \ln\rbracket{1+ e^{-w_k}}.\label{eq:free_energy_density}
\end{equation}
Here, $f\sbracket{w}$ is the free energy density associated to $\rho_w$ and $f\sbracket{w-\eta h}$ is the free energy density associated to $\rho_{\varepsilon_\eta}$.

\subsection{Computation of the charged moments}\label{sec:scgf_at_equilibrium}
Using the results of the previous subsection, we now compute the charged moments
\begin{equation} \label{charged_moments_eqGGE}
    Z^{\text{eq}}_m(\alpha)=\expval{\exp \left(i \textstyle \sum_{p\in F_m} \lambda_{p,\alpha} \, \tilde{N}_{A,p}\right)}_{w^{\otimes m}},
\end{equation}
where $\expval{\ldots}_{w^{\otimes m}}=\tr \rbracket{\ldots\rho_w^{\otimes m}}$ is the expectation value with respect to a multi-copy GGE with generalized Boltzmann weight $w$, generalizing the single-copy definition Eq.~\eqref{expval_w}. 
Notice that due to translational invariance of the state in copy space (i.e., the state is the same in each copy), the Fourier transform $\mathcal{F}_{j \to p}$ preserves factorization, with each 
$p$-mode carrying an independent copy of the same generalized Gibbs ensemble, i.e.,  
\begin{equation}
    \rho_w^{\otimes m} = \otimes_{j=1}^{m} \rho_{w} = \otimes_{p\in F_m}^m \rho_{w}, 
\end{equation}
with $\rho_w$ defined in Eq.~\eqref{expval_w}. Here, $\otimes_{j=1}^m$ refers to the tensor product in the original copy basis, while $\otimes_{p\in F_m}$ denotes the tensor product in the Fourier-transformed copy basis (see Eq.~\eqref{eq:fourier_transform_copy}). This property allows us to factorize the r.h.s. of Eq.~\eqref{charged_moments_eqGGE} as 
\begin{align}
Z^{\text{eq}}_m(\alpha)=\prod_{p\in F_m} \expval{ \exp \left( i \textstyle  \lambda_{p,\alpha} \, \tilde{N}_{A,p}\right)}_{w}.
\end{align}
Furthermore, each expectation value is the FCS of a $U(1)$-charge in a single copy theory, where the results of the previous subsection directly apply.
Therefore, using Eq.~\eqref{eq:scaled_cumulant_generating_function},
we have that 
\begin{equation}
    \prod_{p\in F_m} \expval{ \exp \left(i \textstyle  \lambda_{p,\alpha} \, \tilde{N}_{A,p}\right)}_{w}   
    \asymp 
    \prod_{p\in F_m} \exp \left[-x G(i \lambda_{p,\alpha})\right]=\exp \left[-x \sum_{p\in F_m} G(i \lambda_{p,\alpha})\right],\label{eq:equilibrium_summation_scgf}
\end{equation}
where $x=|A|$ and $G$ is defined in Eq.~\eqref{def:limit_SCGF}. 
Notice that while in Sec.~\ref{sec:large_deviation_theory_flow_equation} $\eta$ was assumed to be real, here we need to make an analytic continuation to complex values. For complex $\eta$, the biased state $\rho_{\varepsilon_\eta}$ does not correspond to a standard MES, so the validity of the thermodynamic arguments is not guaranteed. 
Nonetheless, the analytic continuation of  $\eta$ is possible as long as the SCGF is finite for any real $\eta$~\cite{Characteristic_functions}. Therefore, we can still use Eq.~\eqref{SCGF} together with Eq.~\eqref{eq:free_energy_density}, with the single particle eigenvalue associated to the particle number operator being $h(k)=1$, to get
\begin{equation}
    G(i \lambda_{p,\alpha}) = -\int_{}^{} \ln\rbracket{\frac{1+ e^{-w_k+i \lambda_{p, \alpha}}}{1+ e^{-w_k}}} \frac{\dd k}{2\pi}.
\end{equation}
Performing the summation over $p$ on the r.h.s. of Eq.~\eqref{eq:equilibrium_summation_scgf}, we obtain the asymptotic results for the charged moments at equilibrium
\begin{equation}\label{eq:summation_over_p}
    \ln Z^{\text{eq}}_m(\alpha)
    = x  \int_{}^{}\frac{\dd k}{2\pi} H^{\alpha}_m(k),\quad x=|A|,
\end{equation}
where 
\begin{equation}\label{eq:H_wk}
    H^{\alpha}_{m}(k)=\ln\rbracket{1+e^{-m w_k}e^{i\alpha}}-m\ln\rbracket{1+e^{-w_k}},
\end{equation}
which can be re-written in terms of the density $n_k$ in Eq.~\eqref{eq:n_k_GGE} as
\begin{equation}\label{eq:H_nk}
    H^{\alpha}_{m}(k)=\ln\left[n_k^m e^{i\alpha} + (1-n_k)^m\right].
\end{equation}
See Appendix~\ref{sec:calculation_of_SCGF} for a detailed derivation. 
Eq.~\eqref{eq:summation_over_p} together with Eq.~\eqref{eq:H_wk} or, equivalently, Eq.~\eqref{eq:H_nk} is our main result for the equilibrium part.

Note that for a generic conserved quantity that can be decomposed in the number operator basis $\hat{O}=\sum_{k}h_kN_{k}$, we expect the more general result where the entropy function is modified to take into account the $k$-dependence of the eigenvalue of the charge
\begin{equation}\label{eq:H_hk}
    \tilde{H}^{\alpha}_{m}(k)=\ln\left[n_k^m e^{i\alpha h_{k}} + (1-n_k)^m\right].
\end{equation}
However, note that this formula is specific to observables that admit this decomposition, and, more generally, our derivation is not expected to work for FCS of non-conserved quantities. For instance, in the transverse field Ising model \cite{Groha2018}, the authors study the FCS of the transverse magnetization, which is not a conserved order parameter (this is due to the fact, that in momentum space, the transverse magnetization counts pair creation/annihilation of particles, not number of particles), so that the corresponding charge cannot be written in the number operator basis. Similarly, work statistics is another type of FCS, where the charge is replaced by the work done during the quench. In this case, the work cannot be written in the number operator basis, so that its FCS takes a different form \cite{Perfetto2019}, as already mentioned in \cite{Parez2021}.
\section{Charged moments Out-of-Equilibrium \label{out_of_equilibrium}}
In the previous section, we evaluated the charged moments in a GGE for an asymptotically large subsystem $A$, where the large-deviation principle applies and the corresponding large-deviation function is fixed by thermodynamics or, equivalently, by the equilibrium limit of BFT. In this section, we address the out-of-equilibrium case by mapping the problem to one again amenable to BFT \cite{Myers_Doyon_19}, which in practice amounts to relating the charged moments to the full counting statistics of time-integrated currents.

In Sec.~\ref{sec:ballistic_fluctuation_theory}, we summarize the main features of BFT relevant for our purposes, focusing on time-integrated currents and free fermions.
In Sec.~\ref{symmetry_preserving_states}, we introduce a class of initial states that preserve the $U(1)$ symmetry and emit pairs of ballistically propagating quasiparticles. In Sec.~\ref{sec:quantum_quench}, we give the form of the GGE for this class of states. In Sec.~\ref{evaluation_out_of_eq}, we proceed to the evaluation of the charged moments out-of-equilibrium. In Sec.~\ref{non-asymptotic}, we treat the evalutation of the charged moments in the non-asymptotic regime. 
\subsection{
FCS and SCGF of time-integrated currents \label{sec:ballistic_fluctuation_theory}}
In the following, all operators are understood to be in the Heisenberg picture and therefore acquire an explicit time dependence.
To perform the out-of-equilibrium calculation of the charged moments, we will need to compute the FCS of time-integrated currents. Consider a conserved quantity $Q$ with charge density $q(x,t)$, whose time evolution is governed by the continuity equation
\begin{equation}
    \partial_tq(x,t)+\partial_x j(x,t)=0.
\end{equation}
where $j(x,t)$ denotes the corresponding current density.
The time-integrated current associated to a conserved charge $Q$ in some time interval $\sbracket{0,t}$ at position $x$ is defined by 
\begin{equation}
    J_{x}|_{0}^{t}=\int_{0}^{t} j(x,\tau) \dd \tau,
\end{equation}
and BFT~\cite{Myers_Doyon_19} provides an expression for its FCS, still in a MES, for systems that admit ballistic transport of conserved quantities. 
Assuming large-deviation principle with linear scaling, the SCGF associated to $J_{x}|_{0}^t$ in $\rho_w$ is defined as
\begin{equation} \label{G_dyn_lambda}
    G^\dyn(\eta)=-\lim_{t\to\infty}\frac{1}{t}\ln \expval{e^{\eta J_x|_{0}^{t}}}_w,
\end{equation}
in analogy with Eq~\eqref{def:limit_SCGF}, and fully characterizes the fluctuations of $J_{x}|_{0}^{t}$.

Recall that in Sec.~\ref{sec:large_deviation_theory_flow_equation} we considered a biased GGE $\rho_{\varepsilon_\eta}$, cf. Eq.~\eqref{eq:equilibrium_biased_gge}. In this section we consider its dynamical analogue $\rho_{\varepsilon^\dyn_\eta}$,
where for free fermionic systems the generalized Boltzmann weight $\varepsilon^\dyn_{\eta}$ takes now the form
\begin{equation}
    \varepsilon^\dyn_{\eta}(k)=w_k-\eta h(k)\sgn[v_k]. \label{eq:flow_equation}
\end{equation}
Here $v_k=\dv{E}{k}$ is the dispersion relation and $h(k)$ the single particle eigenvalue associated with $Q$. 
Then BFT~\cite{Myers_Doyon_19}, combined with the theory of generalized hydrodynamics~\cite{Castro-Alvaredo_Doyon_Yoshimura_16, Bertini_Collura_De-Nardis_Fagotti_16}, 
yields the SCGF in the form
\begin{equation}
G^\dyn(\eta)=f^\dyn\sbracket{\varepsilon^\dyn_\eta}-f^\dyn\sbracket{w},\label{SCGF_dyn}
\end{equation}
where
\begin{equation}
\label{f_dyn}
    f^\dyn\sbracket{w}=-\int_{}^{}\frac{\dd k}{2\pi}|v_k|\ln\rbracket{1+e^{-w_k}}.
\end{equation}
\subsection{Quench protocol in the symmetry preserving states \label{symmetry_preserving_states}}
In this section we specify the quench-protocol, by specifing initial states and hamiltonian governing the evolution. We focus on a class of integrable initial states that act as a source of pairs of entangled quasi-particles. This property is crucial, as it allows us to exploit contour-deformation technique first introduced in~\cite{delVecchio_Doyon_Ruggiero_24} and used below in Sec. \ref{evaluation_out_of_eq}. In addition, since we are interested in symmetry-resolved quantities, the initial state must be an eigenstate of the charge, in particular the total particle number operator $N$. These requirements naturally lead to consider a class of two-site translationally invariant states defined for example in~\cite{Bertini2018}.
This class is a two-parameter generalization of the Néel and Dimer states, including the latter as special cases. 

With a slight abuse of notation, in this section we use the same symbol  $\psi$ for fields defined on the lattice, so that $\psi(x=j)$ is only defined at discrete positions $j\in\Z$.
With this convention, the initial state reads
\begin{equation}\label{eq:initial_state}
    \ket{\Psi_0}=\prod_{j=1}^{L/2}\sbracket{a_0 \psi^\dagger(2j)+a_1\psi^\dagger(2j-1)}\ket{0},\quad \abs{a_0}^2+\abs{a_1}^2=1, 
\end{equation}
where $L$ is an even integer and denotes the volume. This state is an integrable state according to the definition in~\cite{Piroli_Pozsgay_Vernier_17}. The parameters $a_0$ and $a_1$ admit a simple physical interpretation: $\abs{a_0}^2$($\abs{a_1}^2$) is the probability of finding a particle on an even (odd) lattice site. 
In particular, the Néel state denoted as $|{N }\rangle$ is obtained for $a_{0}=1 $ and $ a_1=0$, and the dimer state, denoted $|D\rangle$ for $a_{0}=-a_{1}=1/\sqrt{2}$, i.e., 
\begin{align}
|N\rangle&=\prod_{j=1}^{L/2}\psi^{\dagger}(2j)|0\rangle,& |D\rangle&=\prod_{j=1}^{L/2}\frac{\psi^{\dagger}(2j)-\psi^{\dagger}(2j-1)}{\sqrt{2}}|0\rangle.
\end{align}

The dynamics is governed by a lattice Hamiltonian, and the prototypical choice is the tight-binding Hamiltonian, which takes the form
\begin{equation}
    H = \sum_{k \in \rm{BZ}} E_k \psi^\dagger(k)\psi(k), \quad E_k = -2\cos(k),
\end{equation}
where the sum is over the Brillouin zone BZ $=\frac{2\pi}{L}\cbracket{\Z\cup \left[ {-L}/{2},L/2\right)}$. States of the form~\eqref{eq:initial_state} can be interpreted as sources of entangled quasiparticle pairs with momenta $({k,k+\pi})$, where $k$ belongs to half of the Brillouin zone, denoted as RBZ.
This can be seen changing the basis from real space to momentum space, where the states take the form
\begin{align}
|\Psi_0\rangle&=\frac{1}{L^{L/4}}{\rm det}_{L/2}\{e^{-i2jk_{a}}\}_{j,a=1,\cdots,L/2,k_{a}\in {\rm RBZ}}\prod_{k_{j}\in {\rm RBZ}}\chi^{\dagger}_{k_{j}}|0\rangle\label{Fourier_transform_2_sites},
\end{align}
with $k_{1}<k_{2}<\cdots<k_{L/2}$, and with
\begin{align}
\chi^{\dagger}_{k}=\left[\left(a_{0}+a_{1}e^{ik}\right)\psi^{\dagger}(k)+\left(a_{0}+a_{1}e^{ik+i\pi}\right)\psi^{\dagger}(k+\pi)\right],
\end{align}
as demonstrated in App.~\ref{Fourier_transform}. In particular the following factorization holds
\begin{equation} \label{factorization_k_kpi}
    \ket{\Psi_0}\propto\prod_{k\in {\rm RBZ}}\ket{\Psi_{k,k+\pi}}.
\end{equation}
In what follows, we again focus on the hydrodynamic limit, in which the lattice fields are approximated by coarse-grained fields depending on continuous spatial variables, as defined in Sec. \ref{sec:model},  
and where sums over space $x$ and momentum $k$ turn into integrals (in particular for $L \to \infty$, the integration range is given by $x \in [-\infty, \infty]$ and $k \in [-\pi ,\pi]$). We are going to keep the notation $\psi(x)$ and $\psi(k)$ for the coarse-grained fields.
\subsection{Convergence to the GGE after the quantum quench \label{sec:quantum_quench}}
We would like to characterise the GGE attained after the quantum quench. Consider a local operator $q_\mu$, that has support on some region $\mathcal{R}$~\footnote{One can also consider semi-local region, i.e. region with infinite supports as long as the observable $q_\mu$ has an "envelope" that decays sufficiently fast~\cite{Doyon_20}}. In the long time limit, we expect the region $\mathcal{R}$ to relax to some stationary state, in particular one finds that expectation values of local and semi-local operators can be described by a GGE~\cite{Essler_Fagotti_16}. The form of the Generalized Boltzmann weight $w_k$ of the GGE is fixed by ensuring that expectation values of conserved quantities at time $t=0$ are equal to the one with respect to the GGE
\begin{equation} \label{n_k_in_states}
    \tr \left( \rho_w q_\mu(x,0) \right) = \expval{q_\mu(x,0)}{\Psi_0}.
\end{equation}
The associated physical interpretation corresponds to the thermalization of region $\mathcal{R}$ in the large time limit and the reaching of a maximal entropy state, and that the complement of $\mathcal{R}$ acts as a thermal bath. Here, we emphasize the importance of $\mathcal{R}$ to be a local region in space, as expectation values of operators that act on the whole real space will not necessarily reach a steady state~\cite{Essler_Fagotti_16}. In particular, for the state $\ket{\Psi_0}$, the occupation function is  
\begin{align}
n_k&=\frac{1}{2}+|a_{0}a^{*}_{1}|\cos[k-{\rm arg}(a_{0}a^{*}_{1})]. \label{occ_number_2sites}
\end{align}
%
One can then obtain the Boltzmann weight associated to the GGE $w_k$ by inverting Eq.~\eqref{eq:n_k_GGE}. In particular, for the limit cases of the Néel state, denoted with superscript ${\rm N}$ and the dimer state, with superscript ${\rm D}$
\begin{align}
n^{\text{N}}_k=\frac{1}{2}, \qquad 
n^{\text{D}}_k=\frac{1}{2}[1+\cos(k)].
\end{align}
\subsection{Evaluation of the charged moments \label{evaluation_out_of_eq}}
\begin{figure}[H]
\centering

\begin{minipage}{0.48\textwidth}
\centering
\begin{tikzpicture}[>=Stealth]
\draw[->, thick] (0,0) -- (5,0) node[right]{$x$};
\draw[->, thick] (0,1) -- (0,2) node[above]{$t$};

\node[left] at (0,0) {$(0,0)$};
\node[left] at (0,1) {$(0,t)$};
\node[above] at (4.5,1) {$(x,t)$};
\node[below] at (4.5,0) {$(x,0)$};

\draw[red,thick] (0,1) -- (4.5,1); 
\fill[red] (0,1) circle (2pt);
\fill[red] (4.5,1) circle (2pt);
\node[red] at (2.25,1.2) {$A$};

\draw[blue,thick] (0,0.94) -- (0,0);
\draw[blue,thick] (0,0) -- (4.5,0);
\draw[blue,thick] (4.5,0) -- (4.5,0.94);
\node[blue] at (2.25,-0.3) {$\tilde{A}$};

\coordinate (A) at (2.25,0);
\draw[dashed, teal, thick, ->] (A) -- (1.0,1);
\draw[dashed, teal, thick, ->] (A) -- (3.5,1);

\node[below left, teal] at (1.5,0.7) {$-v_k$};
\node[below right, teal] at (3.0,0.7) {$v_k$};
\end{tikzpicture}
\end{minipage}
\hfill
\begin{minipage}{0.48\textwidth}
\centering
\begin{tikzpicture}[>=Stealth]
\draw[->, thick] (0,0) -- (4,0) node[right]{$x$};
\draw[->, thick] (0,2) -- (0,4) node[above]{$t$};

\node[left] at (0,0) {$(0,0)$};
\node[left] at (0,3) {$(0,t)$};
\node[right] at (3,3) {$(x,t)$};
\node[below] at (3,0) {$(x,0)$};

\draw[red,thick] (0,3) -- (3,3); 
\fill[red] (0,3) circle (2pt);
\fill[red] (3,3) circle (2pt);
\node[red] at (1.5,3.2) {$A$};

\draw[blue,thick] (0,2.94) -- (0,0);
\draw[blue,thick] (0,0) -- (3,0);
\draw[blue,thick] (3,0) -- (3,2.94);
\node[blue] at (1.5,-0.3) {$\tilde{A}$};

\coordinate (A) at (1.5,0);
\draw[dashed, teal, thick, ->] (A) -- (0,2.5);
\draw[dashed, teal, thick, ->] (A) -- (3,2.5);

\node[below left, teal] at (1.0,1.0) {$-v_k$};
\node[below right, teal] at (2.1,1.0) {$v_k$};
\end{tikzpicture}
\end{minipage}
\caption{The region $A$ is denoted by the red line, and is equivalent to the straight region from $(0,t)\to (x,t)$. (Left) In the regime $ t \ll x$ there are long range correlations in $A$ due to the propagation of pairs from the initial state. The long range correlations cannot be described by a GGE, hence expectation values along $A$ are not equivalent to the expectation values in a GGE. One can avoid the long range correlation by integrating along $\tilde{A}$ which is defined by the blue line and still connects the point $(0,t)$ to $(x,t)$, but now via the path $(0,t)\to(0,0)\to(x,0)\to(x,t)$. (Right) In the regime $t\gg x$, the long range correlations exist within $\tilde{A}$ but not $A$, and this is what allows the expectation value to be mapped to a GGE. Note that pairs of particle propagate with momentum $k$ and $k+\pi$, the velocity associated to the particle with momentum $k+\pi$ is given by a sine function, hence it is equivalent to $-v_k$.}
\label{fig:A_tildeA_short_time}
\end{figure}
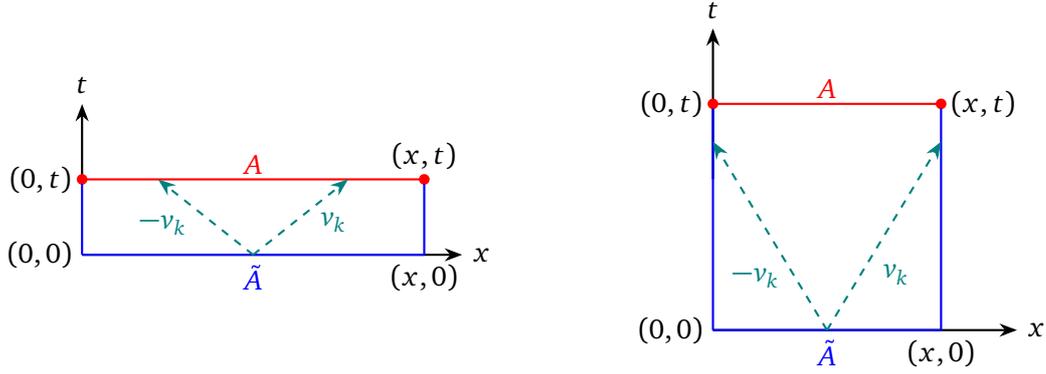

In this section, we compute the charged moments $Z_m(\alpha)$ for the subsystem $A=[0,x]$, after the quantum quench introduced in Sec.\ref{symmetry_preserving_states}, in the ballistic HD regime. 
In particular, here we focus on asymptotic results for 
\begin{equation} \label{Z_m_alpha_ooe}
   Z_{m}(\alpha)=\expval{T_{m,\alpha}(0,t)\bar{T}_{m,\alpha}(x,t)}{\Psi_0^m},
\end{equation}
in the two limits $t\ll x$ and $t\gg x$. Notice that now the composite twist fields pick up a time dependence through the field $\psi$ which in the Heisenberg picture takes the form
\begin{equation}
    \psi(x,t)=\int_{}^{}\frac{\dd k}{2\pi} e^{i(kx-E_kt)}\psi(k).
\end{equation}
We rewrite Eq. \eqref{moments_FCSs} as
\begin{equation}
\expval{T_{m,\alpha}(0,t)\bar{T}_{m,\alpha}(x,t)}{\Psi_0^m}
= \prod_{p\in F_m}
\expval{ e^{i \lambda_{p,\alpha}
N_{t,p}|_{0}^{x}
}
}{\Psi_0},\label{eq:expectation}
\end{equation}
where 
%
\begin{equation}
    N_{t,p}|_{0}^{x}=\int_{0}^{x} \tilde{n}_{p}(x',t) \dd x', \qquad \tilde{n}_{p}(x',t)=\tilde{\psi}^{\dagger}_{p}(x',t)\tilde{\psi}_{p}(x',t)
\end{equation}
In the regime $t\gg x $, the system has equilibrated. As a result, the expectation value of $N_{t,p}|_{0}^x$ is equivalent to the expectation value in a GGE, and we recover the equilibrium result from Sec.~\ref{equilibrium_case}, namely Eq.~\eqref{eq:summation_over_p} together with Eq.~\eqref{eq:H_nk}.
However, for GGEs fixed by the occupation function Eq.~\eqref{occ_number_2sites}, that result can be further simplified (see Appendix~\ref{equiv_results}), so that, eventually, the charged moments read
\begin{equation} 
    \ln Z_m(\alpha)=\frac{i\alpha x}{2}+x \int_{-\pi}^{\pi} \frac{\dd k}{2\pi} \Re H_m^\alpha(k).\label{res_larget}
\end{equation}
On the other hand, in the opposite regime $ t \ll x$, there are correlations within $A$ due to the pair productions in the initial state which need to be taken into account, as illustrated in Fig.~\ref{fig:A_tildeA_short_time}. For operators supported on $A$, expectation values with respect to the initial state $\ket{\Psi_0}$ are not equivalent to expectation values with respect to a GGE. Using a technique of contour deformation, as first introduced in \cite{delVecchio_Doyon_Ruggiero_24}, one can avoid these long-range correlations, thus enabling us to still make use of BFT, which can only be applied along paths with fast-decaying correlations \cite{Myers_Doyon_19,Myers_Bhaseen_Harris_Doyon_20}.
The contour deformation shown in Fig.~\ref{fig:A_tildeA_short_time} follows directly from the continuity equation (by integrating it over the spatial interval $[0,x]$ and temporal interval $[0,t]$), and allows to write
\begin{align}
\int_{0}^{x}\tilde{n}_{p}(x',t)dx'&=-\int_{0}^{t}\tilde{j}_{p}(x,t')dt'+\int_{0}^{x}\tilde{n}_{p}(x',0)dx'-\int_{t}^{0}\tilde{j}_{p}(0,t')dt', \label{eq_decomposition}
\end{align}
where $\tilde{j}_p$ denotes the current density associated to the number operator. Hence, Eq.~\eqref{eq:expectation} can be rewritten as
\begin{equation}
    \expval{e^{i \lambda_{p,\alpha}N_{t,p}|_{0}^{x}}}{\Psi}=\expval{e^{i \lambda_{p,\alpha} \left(-J_{x,p}|_{0}^{t}+N_{0,p}|_{0}^{x}-J_{0,p}|_{t}^{0}\right)}}{\Psi_0},\label{initial_expression}
\end{equation}
where we have used the compact notation
\begin{equation}
    J_{x,p}|_{0}^{t}=\int_{0}^{t} \tilde{j}_{p}(x,t') \dd t'.
\end{equation}
for the integrated current.
We then multiply and divide the r.h.s. of \eqref{initial_expression} by $\langle e^{i \lambda_{p,\alpha} N_{0,p}|_{0}^{x}}\rangle$, and use the result of~\cite{Horvath2025,Horvath2025Scipost} to get
\begin{equation}
    \ln \expval{e^{i \lambda_{p,\alpha}N_{t,p}|_0^x}}=xI_{\text{eq}}(i \lambda_{p,\alpha})+o(x)+tI_{\dyn}(i \lambda_{p,\alpha})+o(t),
\end{equation}
where 
\begin{equation}\label{eq:geq_definition}
    I_{\text{eq}}( i \lambda_{p,\alpha})=\lim_{x\to\infty}\frac{1}{x}\ln \langle \Psi_0 |e^{i \lambda_{p,\alpha}N_{0,p}|_{0}^{x}}|\Psi_0\rangle,
\end{equation}
and 
\begin{equation} \label{eq:I_dyn}
    I_{\dyn}( i\lambda_{p,\alpha})=\lim_{t\to\infty}\frac{1}{t}\lim_{x\to\infty}\ln \frac{\expval{
    e^{-i \lambda_{p,\alpha} J_{x,p}|_{0}^{t}}
    e^{-i \lambda_{p,\alpha} J_{0,p}|_{t}^{0}}
    e^{i \lambda_{p,\alpha}N_{0,p}|_{0}^{x}}}{\Psi_0}}{\langle \Psi_0 |e^{i \lambda_{p,\alpha}N_{0,p}|_{0}^{x}}|\Psi_0\rangle}.
\end{equation}
We further notice that $\ket{\Psi_0}$ in \eqref{eq:initial_state} is an eigenstate for the restricted charge $N_{0,p}|_{0}^{x}$ up to terms of order $1$ for $x$ large, i.e., $\expval{N_{0,p}|_0^x}= x/2+O(1)$. Hence the SCGF \eqref{eq:geq_definition} can be obtained directly from the average value, as
\begin{equation}
    I_{\text{eq}}(i\lambda_{p,\alpha})=\lim_{x\to\infty} \frac{i\lambda_{p,\alpha}}{x}
    \expval{N_{0,p}|_0^x}=\frac{i\lambda_{p,\alpha}}{2},
\end{equation}
Furthermore, this observation, still using results in~\cite{Horvath2025,Horvath2025Scipost}, can be used to rearrange and simplify Eq. \eqref{eq:I_dyn} to give
\begin{equation}
    I_\dyn( i\lambda_{p,\alpha})=\lim_{t\to\infty}\frac{1}{t}\lim_{x\to\infty} \ln \langle \Psi_0| e^{i \lambda_{p,\alpha}\sbracket{-J_{x,p}|_{0}^{t}-J_{0,p}|_{t}^{0}}}|\Psi_0\rangle.
\end{equation}
Then reying on clustering of connected correlation functions at large distance, 
one can split the current contributions, leading to 
\begin{align} \label{two_vertical_paths}
    \lim_{x \to \infty}
    \expval{e^{i \lambda_{p,\alpha} \sbracket{-J_{x,p}|_{0}^{t}-J_{0,p}|_{t}^{0}}}}{\Psi_0}=\langle \Psi_0|e^{-i \lambda_{p,\alpha}J_{0,p}|_{0}^{t}}|\Psi_0\rangle\langle \Psi_0|e^{-i \lambda_{p,\alpha} J_{0,p}|_{t}^{0}}|\Psi_0\rangle
    &=\Big| 
    \langle \Psi_0|e^{i \lambda_{p,\alpha}J_{0}|_{0}^{t}}|\Psi_0\rangle \Big|^2.
\end{align}
Finally 
\begin{equation}
I_\dyn( i  \lambda_{p,\alpha})=\lim_{t\to\infty}\frac{2}{t} \ln | \expval{e^{i \lambda_{p,\alpha}J_{0}|_{0}^{t}}}{\Psi_0} | =\Re\lim_{t\to\infty}\frac{2}{t}\ln \expval{e^{i \lambda_{p,\alpha}J_{0}|_{0}^{t}}}{\Psi_0}.
\end{equation}
Now, on the two vertical contributions of the deformed contour there are no correlations left, and in the limit $t\to\infty$, expectation values of local observables such as $J_{0}|_{0}^{t}$ can be substituted with expectation values in a GGE (cf. Eqs.~\eqref{G_dyn_lambda} and~\eqref{SCGF_dyn}) leading to
\begin{equation} \label{Idyn_Gdyn}
    I_\dyn(i\lambda_{p,\alpha}) = -2\Re G^\dyn(i\lambda_{p,\alpha}) \ .
\end{equation}
Then, one can use the BFT result, Eq.~\eqref{SCGF_dyn} together with Eqs. \eqref{f_dyn} and \eqref{eq:flow_equation}, with single particle eigenvalue $h(k)=1$, to evaluate the r.h.s. of Eq.~\eqref{Idyn_Gdyn}.

Finally, to obtain the charged moments one can perform the summation over $\lambda_{p,\alpha}$,
\begin{align}
Z_{m}(\alpha)=\prod_{ p\in F_m}
\expval{ e^{i \lambda_{p,\alpha} N_{t,p}|_{0}^{x}}}{\Psi_0} =e^{\sum_{p}\frac{i \lambda_{p,\alpha}x}{2}-2t\Re\left[\sum_{p}G^{\rm dyn}( \lambda_{p,\alpha})\right]}
             =e^{\frac{i\alpha x}{2}-2t\Re\left[\sum_{p}G^{\rm dyn}( \lambda_{p,\alpha})\right]} \ .
\end{align}
After taking the logarithm 
\begin{equation}
    \ln Z_m(\alpha)=\frac{i\alpha x}{2}+2t\Re\int_{}^{}  \frac{\dd k}{2\pi} |v_k|\sum_{p\in F_m}\sbracket{ \ln (1+e^{-w_k+i \lambda_{p,\alpha}\, \sgn\sbracket{v_k}})-\ln \rbracket{1+e^{-w_k}}} ,\label{log_Zalpha}
\end{equation}
and performing the sum over $p$ with the explicit form of $\lambda_{p,\alpha}$ (similarly to what was done in Sec.~\ref{sec:scgf_at_equilibrium}), we obtain
\begin{equation}
    \ln Z_m(\alpha)=\frac{i\alpha x}{2}+2t\int_{}^{}  \frac{\dd k}{2\pi}|v_k| \Re \sbracket{\ln\rbracket{1+e^{-mw_k}e^{i\alpha\,\sgn [v_k]}}-m\ln \rbracket{1+e^{-w_k}}}. \label{eq:intermediate_res}
\end{equation}
Furthermore, the integrand is the real part of a logarithm.
As the complex dependence comes from the factor $i\alpha$ inside of the absolute value, the sign of $\alpha$ does not matter in the final expression, so that one can get rid of the sign dependence $\sgn[v_{k}]$. As a consequence, Eq.\eqref{eq:intermediate_res} can be further simplified in terms of $H_{m}^\alpha (k)$ (c.f. Eq.~\eqref{eq:H_wk}) and reads
\begin{equation}
    \ln Z_m(\alpha)=\frac{i\alpha x}{2}+2t \int_{}^{}  \frac{\dd k}{2\pi} \abs{v_k} \Re H_{m}^{\alpha} (k) . \label{eq:final_result}
\end{equation}
Putting everything together, our result for the out-of-equilibrium result in the two asymptotic regimes can be summarized as 
\begin{equation}\label{result_parez}
   \ln Z_m(\alpha)=\begin{cases}&\frac{i\alpha x}{2}+2t\int_{-\pi}^{\pi} \abs{v_k}  \frac{\dd k}{2\pi} \Re H_{m}^{\alpha}(k)\quad t\ll x\\
       &\frac{i\alpha x}{2}+x\int^{\pi}_{-\pi}  \frac{\dd k}{2\pi} \Re H_{m}^{\alpha}(k)\quad t\gg x
   \end{cases}.
\end{equation}
Eq.~\eqref{result_parez} recovers the result of~\cite{Parez2021} for the N\'eel and dimer states, originally obtained using free-fermion techniques, and extends it to the two-site translationally invariant states defined in Eq.~\eqref{eq:initial_state}.
The result was derived under the assumption of an integrable (pair-producing) initial state that is an eigenstate of the total charge \(N\).
For the class of states in Eq.~\eqref{eq:initial_state}, this further implies that the state is also an eigenstate of the restricted charge \(N_A\), something that also enters in the derivation.

In the long-time regime all the odd cumulants (imaginary part of the charged moments) vanish except for the first one, which is a peculiarity of $n_k$ associated with $\ket{\Psi_0}$ (c.f. Eq.~\eqref{n_k_in_states}).
In the short-time regime the explicit form of $n_k$ is not used, instead. 
In the latter case, this feature appears to be more general, as it follows from the combination of the two vertical-path contributions in Eq.~\eqref{two_vertical_paths}, which enter as complex conjugates. As a consequence, only the even cumulants (real part of the charged moments) receive a contribution from the currents.
This structural property admits a simple physical interpretation.
In integrable, pair-producing initial states, current fluctuations arise from symmetric pairs of excitations.
As a result, odd connected correlators of currents, which are odd under the symmetry exchanging the two partners of a pair, vanish identically.
Since the time-dependent contribution to the charged moments is entirely encoded in the full counting statistics of the integrated current through $I_{\mathrm{dyn}} (\lambda_{p,\alpha})$, 
this directly implies that 
all odd cumulants generated by the expansion of
$\ln Z_m(\alpha)$
do not contribute to the time-dependent part.
\subsection{Non-Asymptotic Regime \label{non-asymptotic}}

The asymptotic results of the previous section can be extended to all rays $\xi = x/t$ within the ballistic regime, by using similar arguments, but now for \emph{single-mode twist fields} $\tau_{\eta,k}(x,t)$ (for mode $k$),
allowing to take into account the meaning of
“short” and “long” time depending on the speed of the traveling particles.
Single mode twist fields were introduced in \cite{delVecchio_Doyon_Ruggiero_24} (see Sec. 4.4) and can be used to factorise any $U(1)$ twist field (cf. \eqref{eq:Twist_varphi_Nx}) as 
\begin{equation}
    T(\eta; x)= \prod_{k} \tau_{\eta,k} (x,t) \ .
\end{equation}
In our case $k \in \rm{RBZ}$.
Moreover, our initial single-copy state factorizes in pairs, according to Eq.~\eqref{factorization_k_kpi}, so the relevant factorization of the twist fields is in \emph{pair-mode twist fields} 
\begin{equation} \label{tau_pair}
    \tau_{\eta, [k] } (x,t) \equiv  \tau_{\eta, k} (x,t) \tau_{\eta, k + \pi} (x,t),
\end{equation}
associated with the pair of momenta $(k, k + \pi)$.

In the replica theory, the two-point correlation function of our charged branch-point twist field admits the following factorization
\begin{equation} \label{double-factorization-twist-fields}
       \expval{T_{m,\alpha}(0,t)\bar{T}_{m,\alpha}(x,t)}{\Psi_0^m} = \prod_{p \in F_m} \prod_{k \in \rm{RBZ}} 
       {}_p\expval{\tau_{\lambda_{p,\alpha}, [k]}(0,t)\bar{\tau}_{\lambda_{p,\alpha}, [k]}(x,t)}{\Psi_{k, k + \pi}}_p.
\end{equation}
While the microscopic structure of our initial states differs from the squeezed (Bogoliubov) states analysed in~\cite{delVecchio_Doyon_Ruggiero_24}, 
the HD analysis based on single-mode and pair-mode twist fields and the associated BFT remains valid, with only minor modifications discussed below.

In the HD limit, the spacetime dependence of correlation functions is governed solely by quasiparticle kinematics.
For a generic dispersion relation $E_{k}$, quasiparticles propagate ballistically with velocity $v_k = \partial_k E_k$.
The admissible integration paths in the BFT construction are determined by the presence/absence of strong correlations in charge and current density correlations. 
In this respect, the presence of saddle points in such two-point correlators  play a key role, as saddle points correspond to kinematic configurations where ballistic correlations are enhanced (see Appendix B.5 in \cite{delVecchio_Doyon_Ruggiero_24}).
In particular,
for two points in spacetime, $(x,t)$ and $(x',t')$, the saddle point condition is of the form
\begin{equation} \label{single-mode-saddle}
\abs{\frac{x-x'}{t-t'}} = v_k \qquad \text{(single-mode contributions)},
\end{equation}
corresponding to one particle propagating between the two spacetime points
, or
\begin{equation} \label{pair-mode-saddle}
\frac{|x-x'|}{t+t'} = v_k \qquad \text{(pair-mode contributions)},
\end{equation}
for pair correlated initial states, corresponding to two distinct particles, created together at the same spacetime point and propagating in opposite directions.

What differs between initial states is the pairing structure of quasiparticle modes entering anomalous correlations.
For squeezed states, one has particle--particle pairing $\langle \psi(k) \psi(-k)  \rangle \neq 0$,
whereas for lattice states such as Néel and dimer states the pairing is of particle--hole type, $\langle \psi^\dagger(k) \psi({k+\pi}) \rangle \neq 0$.
Accordingly, while the relevant pair-mode twist fields couple modes $(k,-k)$ for squeezed states, they couple $(k,k+ \pi)$ for our two-parameter class of states, leading to the factorization \eqref{double-factorization-twist-fields}.

The derivation of the pair-mode contribution proceeds exactly as in~\cite{delVecchio_Doyon_Ruggiero_24} (cf. Eqs.~(108)--(112) there).
The analysis involves two independent kinematic conditions: one ensuring the suppression of genuinely pair-induced contributions associated with the saddle point \eqref{pair-mode-saddle} (allowing replacement of initial-state expectation values by GGE averages), and a second ensuring sufficiently fast decay of charge and current correlations associated with the single-mode saddle point \eqref{single-mode-saddle}, which is present both in the initial state and in the GGE.

In the hydrodynamic limit, each momentum mode contributes independently,
and its contribution saturates once the ballistic separation of the
corresponding quasiparticle pair exceeds the size of the interval.
The final result for the charged moments after the quench defined in~\eqref{Z_m_alpha_ooe} evaluated via~\eqref{double-factorization-twist-fields} reads 
\begin{equation}
\ln Z_m(\alpha)
=
\frac{i\alpha x}{2}
+
\int^{\pi}_{-\pi} \frac{dk}{2\pi} \, 
\min [ 2t |v_k|, x ] \,
{\rm{Re}} \, H_m^{\alpha}(k).
\label{eq:min_formula_lattice}
\end{equation}
Note that, since we are on the lattice (the Brillouin zone is compact and the velocity bounded), the velocity is bounded,
$|v(k)|\le v_{\max}$.
This implies a uniform early-time regime:
for $t<x/(2v_{\max})$ one has $2|v(k)|t<x$ for all $k$, 
corresponding to a linear growth in time.
For later times, $t>x/(2v_{\max})$, only a subset of modes has saturated,
and the integral in~\eqref{eq:min_formula_lattice} naturally splits into
saturated and unsaturated momentum regions, exactly as in the continuum case.

From the correlations viewpoint, in the continuum, where the quasiparticle velocity is monotonic and
unbounded, a saddle point exists for any ray, and the choice of path is
dictated by the requirement of avoiding this coherent contribution.
On the lattice, instead, the boundedness of the velocity implies the
existence of spacetime regions in which no stationary point is present.
In these regimes correlations are strongly suppressed, and the
factorisation underlying BFT applies
automatically.
This difference affects the interpretation of the various spacetime
regimes and the path selection, but does not alter the final hydrodynamic
expression~\eqref{eq:min_formula_lattice}.

Eq.~\eqref{eq:min_formula_lattice} is our final result for charged moments after the quench. It is in agreement with Ref.~\cite{Parez2021} for Neel and Dimer state and extends it to the two-sites translational invariant states defined in~\eqref{eq:initial_state}. 
Instead, it is not expected to hold for more general integrable states~\cite{Piroli_Pozsgay_Vernier_17}, which are not eigenstates of the particle number operator, and therefore not considered here.

\section{Symmetry Resolved Entanglement}\label{sec:symm_res_Fourier_transform}
The symmetry-resolved R\'enyi entropies $S_m(q)$, c.f. Eq.~\eqref{symmetry_resolved}, are obtained via the Fourier transform of the charged moments, Eq.~\eqref{moments_fourier}. In general, it is not possible to obtain exact results for arbitrary $n_k$. Nevertheless, the saddle point approximation can capture the leading term behavior, at least in some cases, as we are now going to see.

Starting from the equilibrium scenario, we would like to compute 
\begin{equation}
    \mathcal{Z}_m(q)= \int_{- \pi}^{\pi} \frac{\dd \alpha}{2\pi}e^{-i\alpha q} Z_m (\alpha) =
    \int_{-\pi}^{\pi} \frac{\dd \alpha}{2\pi} e^{x\Phi(i\alpha)}, \quad \Phi(i\alpha)=-\frac{i\alpha q}{x}+\int_{}^{}  \frac{\dd k}{2\pi}H_m^\alpha(k),
\end{equation}
where we recall that $q$ is the eigenvalue of $N_A$, identifying a specific sector. The saddle-point condition is given by $\partial_{i\alpha}\Phi(i\alpha^*)=0$ and reads
\begin{align}
    \frac{q}{x}= \int_{}^{}  \frac{\dd k}{2\pi} \partial_{i\alpha}H_{m}^{\alpha}(k)|_{\alpha^*}, \qquad \partial_{i\alpha}H_{m}^{\alpha}(k)|_{\alpha^*}= 
    \frac{n_k^m e^{i\alpha^*}}{n_k^m e^{i\alpha^*}+(1-n_k)^m}.
\end{align}
It is in general not possible to determine the saddle-point solution analytically from the above expression. However, for values of $q$ around the mean $\expval{N_A}_m \equiv \partial_{i \alpha} \ln Z_m(\alpha)|_{\alpha=0}=x\int \frac{\dd k}{2\pi}\partial_{i\alpha}H_m^\alpha|_{\alpha=0}$, i.e. for deviations subextensive in $x$,
$\Delta q_m \ll x $, the saddle point lies at $\alpha^*=0$ and we may expand $\Phi(i\alpha)$ to second order, getting 
\begin{align}
    \Phi(i\alpha) \approx \Phi(0)+  i\alpha \Phi'(0) - \frac{\alpha^2}{2}\Phi''(0)
    =-i \alpha\frac{\Delta q_m}{x} -\frac{\alpha^2}{2}\int_{}^{}  \frac{\dd k}{2\pi}\partial_{i\alpha}^2 H_m^\alpha(k)|_{\alpha=0},
\end{align}
where $\Phi(0)=0$ and $\Delta q_m=q - \expval{N_A}_m $. 
Then, the charged moments become Gaussian, and one can perform the Fourier transform explicitly,
\begin{equation}
    \mathcal{Z}_m(q)=\int_{-\pi}^{\pi}  \frac{\dd \alpha}{2\pi} e^{-i\alpha \Delta q_m - \frac{x\alpha^2\Phi''(0)}{2}} \approx \frac{1}{\sqrt{2\pi \sigma_m^2}}\exp \rbracket{-\frac{(\Delta q_m)^2}{2\sigma_m ^2}},
\end{equation}
where $\sigma_m^2 = x\Phi''(0)=x \int_{}^{}\frac{\dd k}{2\pi}\partial_{i\alpha}^2 H_m^{\alpha}|_{\alpha=0}$ is the variance. 
However, it is important to note that while in order to compute the symmetry resolved entropy in Eq.~\eqref{symmetry_resolved}, both $\mathcal{Z}_m (q)$ and $\mathcal{Z}_{1}(q)$ are required, in general, $q$ cannot satisfy simultaneously the two equations $\Delta q_{m}/x\ll 1$ and $\Delta q_{1}/x \ll 1$. Therefore, for generic GGE profiles, the Gaussian approximation does not determine $S_m(q)$ in general.
Only for special occupation numbers such that $\langle N_{A}\rangle_{m}=\langle N_{A}\rangle_{1}$ the Gaussian approximation becomes applicable. %
Note also that the replica limit $m \to 1$ is subtle and not automatically captured by the Gaussian approximation in a generic GGE, and, as a result, the SREE $S(q)$ remains controlled by the large deviation function.

Next, we consider the out-of-equilibrium quench scenario, and we restrict to our two-parameter class of initial states, with occupation number given by Eq.~\eqref{occ_number_2sites}. In this case, the analysis done in~\cite{Parez2021} directly applies. We repeat it here for completeness.
The saddle point equation of Eq.~\eqref{eq:min_formula_lattice} is given by 
\begin{equation} \label{saddle_point_OOE}
-  \Delta q_1
+
\int^{\pi}_{-\pi} \frac{dk}{2\pi} \, 
\min [ 2t |v_k|, x ] \,
\partial_{i \alpha} {\rm{Re}} \, H_m^{\alpha}(k) = 0.
\end{equation}
Here notice that for this particular class of states $\Delta q_m = \Delta q_1=\Delta q$.
Similarly to the equilibrium case it is still not possible to determine an analytical result for the saddle-point $\alpha^{*}$, for all values of the two parameters $a_0$ and $a_1$, except for the case of the Néel state~\cite{Parez2021}. 
However, general features can be inferred without relying on the explicit calculation of the integral.
In particular, one can infer the existence of an initial \emph{time-delay} followed by a growth in time for the symmetry-resolved entanglement entropy. This time delay arises from the bounded group velocities of the free fermions. In fact for the tight-binding model one has $v_{k}\in [-2,2]$, giving a bound on the integral in \eqref{saddle_point_OOE}, which takes values only in $[-4/\pi,4/\pi]$ independently on the specific form of $n_k$ (note that $\partial_{i \alpha} \Re H_m^{\alpha} (k) \in [0,1]$). As a consequence, at early times, no solution to the saddle-point equation exists, resulting in a universal time-delay given by $t_{\rm D} = \frac{\pi}{4} |\Delta q|$ before a saddle-point solution appears.

Similarly to the equilibrium case, the Gaussian expansion remains valid around $\alpha^*=0$, for $\Delta q=q-\frac{x}{2}\ll x$ (i.e., subextensive).
Here, one can use the fact that $\Re H^{\alpha}_{m}(k)$ is an even function of $\alpha$, so that its derivative with respect to $\alpha$ is an odd function of $\alpha$, implying that $\partial_{i\alpha}\Re H^{\alpha}_{m}(k)|_{\alpha=0}=0$. In this regime, a Gaussian approximation is valid again for the charged moments, which can be written as
\begin{equation}
Z_{m}(\alpha; x,t )\approx Z_{m}(0; x, t) e^{-\mathcal{I}_{m}(x,t)\alpha^{2}}, \quad   \mathcal{I}_m (x,t)= \frac{1}{2} \int^{\pi}_{-\pi} \frac{dk}{2\pi} \, 
\min [ 2t |v_k|, x ] \,
\partial^2_{i \alpha} {\rm{Re}} \, H_m^{\alpha}(k)|_{\alpha=0},
\end{equation}
where we made the dependence on $x$ and $t$ explicit.
Their Fourier transform gives a Gaussian distribution in the charge sector
\begin{equation}
\mathcal Z_m(q;x,t)\simeq
Z_m(0;x,t)\,
\frac{1}{\sqrt{4\pi \mathcal I_m(x,t)}}
\exp\!\left[-\frac{(\Delta q)^2}{4\mathcal I_m(x,t)}\right].
\end{equation}
which is valid now for any $m$. Therefore, in this case one can get an analytical expression for the symmetry-resolved Renyi entropies \eqref{symmetry_resolved}
\begin{equation}
S_m(q;x,t)
=
S_m(x,t)
+ C_m(x,t)
-\frac{(\Delta q)^2}{4(1-m)}
\left(
\frac{1}{\mathcal I_m(x,t)}-\frac{m}{\mathcal I_1(x,t)}
\right),\label{symm_resolv_expression}
\end{equation}
with $S_{m}(x,t)$ the total Renyi entropy, and $C_m (x,t)$ is now given by
\begin{equation}
C_m(x,t)=\frac{1}{1-m}
\left[
-\frac12\ln(4\pi \mathcal I_m(x,t))
+\frac m2\ln(4\pi \mathcal I_1(x,t))
\right].
\end{equation}
Now in Eq.~\eqref{symm_resolv_expression} the dependence on $q$ appear only at order $(\Delta q)^{2}$ induced by the charge fluctuations, which is the statement of \emph{equipartition} \cite{Xavier2018,turkeshi2020entanglement,Bonsignori2019,oblak2022equipartition} at leading order. 
Note that in hydrodynamic regimes one has
\begin{equation}
\mathcal I_m(x,t)= O( J(x,t)),
\qquad
J(x,t)=\int dk\,\min [2v_k t, x],
\end{equation}
giving the scaling
\begin{align}
S_m(x,t) = O(J), \quad 
C_m(x,t) = O(\log J), 
\end{align}
and the sector dependent contribution scales as $O \left(
{(\Delta q)^2}/{J} \right)$.
Therefore it is $O(1)$ in the fluctuation regime $\Delta q=O(\sqrt{J})$,
while it becomes $O(1/J)$ in the central-sector regime $\Delta q=O(1)$,
vanishing in the hydrodynamic limit.

Beyond the Gaussian regime, the saddle-point approximation is not analytically tractable. However, it is possible to obtain the Fourier transform of the charged moments in the discrete case, before taking the thermodynamic limit. One can observe that the normalized charged moments can be interpreted as the generating function of a Poisson–Binomial process. With this interpretation, the Fourier transform of the charged moments can be obtained as the probability distribution of a Poisson–Binomial process. See Appendix~\ref{Poisson-Binomial} for details.
\section{Conclusion \label{conclusion}}

In this article, we studied the charged moments of a RDM,
defined in the presence of a global internal symmetry associated with a conserved charge $Q$, in the context of free fermionic theories.
We provided fully analytic expressions for them both
in generic GGEs and after a class of integrable quantum quenches, and finally comment on the implications for the properties of the symmetry resolved entropies, obtained via Fourier transform.
While at equilibrium the analysis is done for generic free fermions models, out of equilibrium we focus on lattice systems, where it is easy to construct states which are both integrable (where our method applies), and which commutes with $Q$ (the charge we want to symmetry-resolve). 
Regarding quenches, our formula reproduces previous results existing in the literature for specific initial states (i.e., Neel and Dimer), which were obtained by exploiting standard free-fermions techniques \cite{Parez2021}.
Our paper provides a new derivation of charged moments based on twist-field correlation functions and hydrodynamic fluctuations, following the original proposal of  \cite{delVecchio_Doyon_Ruggiero_24}, and extending it via the introduction of composite twist-fields. 
It also provides a consistent derivation of the phenomenological results obtained from the quasiparticle picture~\cite{Calabrese2005, Calabrese2018, Parez2021}.

Our result admits a natural reinterpretation in terms of static and dynamical full counting statistics, making explicit the relation between charged moments and fluctuations on a single copy.
At equilibrium, for quadratic theories, it is well known that Renyi entropies (i.e., the $\alpha=0$ case) can be written in terms of the charge (number of particles) cumulants \cite{Klich2009, Song2012,Calabrese_2012}. More recently, this relation was extended by relating Renyi entropies to a dynamical free energy associated to the large-deviation theory for charge transport 
\cite{delVecchio_Doyon_Ruggiero_24} (see also \cite{Jin_2021} for qualitative arguments). It is then immediate to generalize that relation by rewriting our result in Eq.~\eqref{eq:min_formula_lattice} as
\begin{equation}
    \ln Z_m ({\alpha}) = 2t \sum_{p \in F_m} I^{<, \xi} (i \lambda_{p, \alpha}) + x \sum_{p \in F_m} G^{>,\xi} ( i \lambda_{p, \alpha})
\end{equation}
where we defined the static and dynamic full counting statistics in the final GGE for “slow" and “fast" fermionic modes
\begin{equation}
    G^{>,\xi} (i \eta) = \lim_{x \to \infty} x^{-1} \ln \expval{e^{i \eta N_{>}(x)}} \ ,
    \qquad
    I_{\text{dyn}}^{<, \xi} (i \eta)=  \lim_{t \to \infty} t^{-1} \ln \expval{e^{i \eta J_{N_{<}}(t)}}
\end{equation}
where $N_{<,>} = \int_{|v_k| >,< \xi /2} \dd k \, \psi^{\dag}(k) \psi(k)$ is the conserved quantity
giving the total number of fermionic modes  with speed $v_k$ smaller ($<$) or larger ($>$) than $\xi/2$, with $\xi=x/t$. And $J_{N_{<,>}} (t)$ is the current of slow or fast modes passing
through a point in the time interval $[0,t]$.

Related to that, it is interesting to notice that in our final formula~\eqref{eq:min_formula_lattice}
odd cumulants (except the mean) do not appear.
While for generic GGEs, or in the late-time limit for more general states, all cumulants may a priori contribute (cf.~\eqref{eq:summation_over_p}),
the absence of odd cumulants in the time-dependent part of the charged moments turns out to be a more general feature.
Physically, this originates from the pair structure of integrable initial states, which enforces a balanced left/right contribution of current-carrying excitations.

Several future directions naturally follow from our work. A first immediate generalization concerns the extension of our analysis to broader classes of integrable initial states. In particular, it would be interesting to investigate asymmetric integrable states, such as squeezed states, also relevant in the context of entanglement asymmetry~\cite{Ares2023}. 
On one side, our approach suggests that these results do not extend in a straightforward manner. This is because the charged moments considered here are directly related to the FCS of the subsystem charge after a quench and, in this respect, previous studies of charge fluctuations starting from squeezed initial states~\cite{Horvath2025,Horvath2025Scipost} (see also \cite{Bertini_2024}) have revealed a more intricate post-quench dynamics.
At the same time, possible extensions in this direction should be understood in comparison with those of Refs.~\cite{rylands2024microscopic,ares2023lack,Ares2023}, based on exact free-fermion calculations and on a generalized quasiparticle picture.
Moreover, within the realm of integrable quenches, another possible extension concerns the recently studied class of cross-cap states~\cite{caetano2022crosscap}, which have also been shown to obey a generalized form of the quasiparticle picture~\cite{chalas2024,chalas2025}.

Another natural extension of our framework concerns free bosonic theories. The methods developed here apply straightforwardly to systems with quadratic bosonic Hamiltonians, and the simplest interesting example would be a complex bosonic field theory with a global $U(1)$ symmetry (symmetry resolved entaglement for such model was investigated, e.g., in \cite{Murciano_Ruggiero_2020,Murciano_DiGiulio_2020}). We expect that this setting will provide a useful testing ground for symmetry-resolved entanglement in bosonic systems.

Further directions include the study of more general charged moments involving charges defined on multiple spatial regions \cite{Parez_Bonsignori_Calabrese_Long,gaur2024multi,Bertini_2024}. 
This is relevant for multipartite entanglement measures~\cite{Alba2019Europhys}.
Understanding how such multi-charge observables fit within the present hydrodynamic and twist-field framework would also connect naturally to recent developments on measurements \cite{oshima2023charge,Travaglino_2025}.

It would also be interesting to extend the present analysis beyond homogeneous quantum quenches and beyond the simple class of particle-pair initial states~\cite{Bertini2018}. More general inhomogeneous or slowly varying initial conditions, as well as states with long-range correlations, fall outside the regime where the present simplifications apply. These situations could potentially be addressed within the recently developed ballistic macroscopic fluctuation theory (BMFT)~\cite{Doyon2023}, which provides a more general framework not relying on a GGE flow and is able to capture genuinely long-range correlations.

Finally, an important open challenge is the extension of these ideas to interacting integrable models, and ultimately beyond integrability. In interacting systems, the structure of charged moments and their relation to full counting statistics is expected to be considerably richer, due to mode-dependent dressing effects and nontrivial correlations. Understanding whether an effective hydrodynamic description of charged moments survives in this setting, and how it may cross over to diffusive or chaotic regimes, remains an interesting direction for future investigation.

\section*{Acknowledgements}
PR thanks Pasquale Calabrese and his group at SISSA for useful discussions. The authors thank Benjamin Doyon and Dávid X. Horváth for discussions and collaborations on closely related topics. 
%
\paragraph{Funding information}
P.R. acknowledges support from the UK Engineering and Physical Sciences Research Council (EPSRC) through a New Investigator Award, grant number EP/Y015363/1.

\begin{appendix}
\numberwithin{equation}{section}
\section{Diagonalization of the Branch-point Twist Fields}\label{sec:diagonalization_branch_point_twist_field}
In this section we show that the branch-point twist field $T_m$ is diagonalized by a discrete Fourier transform, Eq.~\eqref{eq:twist_field_on_fourier_basis}. 
\begin{align}
    T_m\tilde{\psi}_p  &=\frac{1}{\sqrt{m}}\sum_{j=1}^m  e^{i\lambda_p j}\psi_{j+1}T_m,\\
                    &=e^{-i\lambda_p}\frac{1}{\sqrt{m}}\sum_{j=1}^m e^{i\lambda_p(j+1)}\psi_{j+1}T_m,\\
                    &=e^{-i\lambda_p}\frac{1}{\sqrt{m}}\rbracket{\sum_{j=1}^{m-1}e^{i\lambda_p(j+1)}\psi_{j+1}+e^{i\lambda_p(m+1)}\psi_{m+1}}T_m,\\
                    &=e^{-i\lambda_p}\frac{1}{\sqrt{m}}\rbracket{\sum_{j=2}^m e^{i\lambda_p j}\psi_j-\psi_1 e^{i\lambda_p(m+1)}}T_m,
\end{align}
for $\tilde{\psi}_p$ to be an eigenfunction of $T_m$ we require that $e^{i\lambda_p m}=-1$. One needs to obtain the $m$ distinct roots for this equation, and select nondegenerate values of $\lambda_{p}$. For fermions 
\begin{equation}\label{eq:fermionic_kp}
    \lambda_p = \frac{\pi(2p-1)}{m},\quad p\in F_{m}=\cbracket{-\frac{m}{2}+1,-\frac{m}{2}+2,\ldots,\frac{m}{2}}.
\end{equation}

Thus, with the appropriate choice of $\lambda_p$, given by Eq.~\eqref{eq:fermionic_kp}, $\tilde{\psi}_p$ diagonalizes the branch-point twist field $T_m$ with eigenvalue $e^{-i\lambda_p}$. The distribution of $\lambda_{p}$ is symmetric, such that their sum is zero
\begin{equation}
    \sum_{p\in F_m}\lambda_p = 0 .
\end{equation}
\section{Detailed calculation for the SCGF}\label{sec:calculation_of_SCGF}
In this section the summation in Eq.~\eqref{eq:equilibrium_summation_scgf} is performed explicitly and we show that 
\begin{equation}
    \sum_{p\in F_m}G(i\lambda_{p,\alpha})=\int_{}^{}\frac{\dd k}{2\pi}H_m^\alpha(k).
\end{equation}
More specifically we would like to show that 
\begin{equation}\label{eq:summation_p_trivial}
    \sum_{p\in F_m}\int_{}^{}\frac{\dd k}{2\pi} \left[\ln\rbracket{1+e^{-w_k}e^{i\lambda_{p,\alpha}}}-\ln\rbracket{1+e^{-w_k}} \right]= \int_{}^{}\frac{\dd k}{2\pi} \left[\ln\rbracket{1+e^{-mw_k}e^{i\alpha}}-m\ln \rbracket{1+e^{-w_k}} \right]
\end{equation}
where
\begin{equation}
    \lambda_{p,\alpha}=\frac{\pi(2p-1)+\alpha}{m}, \quad F_m=\cbracket{\frac{-m}{2}+1,\frac{-m}{2}+2,\ldots,\frac{m}{2}}.
\end{equation}
First note that the term that does not depend on $p$ is trivially computed
\begin{align}
\sum_{p\in F_m}\int\frac{\dd k}{2\pi}\ln\left(1+e^{-w_k}\right)\frac{dk}{2\pi}=m\int\frac{\dd k}{2\pi}\ln\left(1+e^{-w_k}\right).
\end{align}
For the $p$ dependent part we use the Taylor expansion $\ln(1+x)=\sum_{r=1}^\infty \frac{(-1)^{r+1}}{r}x^r$, such that
\begin{align}
    \sum_{p\in F_m}\ln(1+e^{-w_k}e^{i\lambda_{p,\alpha}})&=\sum_{p=-\frac{m}{2}+1}^{m}\ln\rbracket{1+e^{-w_k+i\rbracket{\alpha-\pi}/m}e^{2\pi i p/m}}\\
                                                         &=\sum_{p=-\frac{m}{2}+1}^{\frac{m}{2}}\sum_{r=1}^\infty \frac{(-1)^{r+1}}{r}e^{-rw_k}e^{ir(\alpha-\pi)/m}e^{2\pi i pr/m}\\
                                                         &=\sum_{r=1}^\infty \frac{(-1)^{r+1}}{r}e^{-rw_k}e^{ir(\alpha-\pi)/m} \frac{e^{2\pi i r/m}\rbracket{e^{-\pi i r}-e^{\pi i r}}}{1-e^{2\pi i r/m}},\label{line:summation_p}
\end{align}
where on the last line we have explicitly performed the summation over $p$ using the geometric series result
\begin{equation}\label{eq:summation_p_explicit}
    \sum_{p=-\frac{m}{2}+1}^{\frac{m}{2}}e^{2\pi i pr/m}=\frac{e^{2\pi i r/m}\rbracket{e^{-\pi i r}-e^{\pi i r}}}{1-e^{2\pi i r/m}}.
\end{equation}
For any $r\in\mathbb{N}$ that is not a multiple of $m$ we have that $e^{-\pi i r}-e^{\pi i r}=0$. For $r=sm$, where $s\in\mathbb{N}$, the expression is non zero, taking the limit $r=sm$ we have that
\begin{equation}
    \frac{e^{2\pi i r/m}\rbracket{e^{-\pi i r}-e^{\pi i r}}}{1-e^{2\pi i r/m}}=m\sum_{s=1}^{\infty}\delta_{r,sm}.
\end{equation}
Substituting this in Eq.~\eqref{line:summation_p} we have that
\begin{align}
    &\sum_{r=1}^\infty \frac{(-1)^{r+1}}{r}e^{-rw}e^{ir(\alpha-\pi)/m} \frac{e^{2\pi i r/m}\rbracket{e^{-\pi i r}-e^{\pi i r}}}{1-e^{2\pi i r/m}}\\
    &=\sum_{s=1}^\infty \frac{(-1)^{sm+1}}{sm}e^{-smw}e^{ism(\alpha-\pi)/m} \frac{e^{2\pi i (sm)/m}\rbracket{e^{-\pi i sm}-e^{\pi i sm}}}{1-e^{2\pi i (sm)/m}}\\
    &=\sum_{s=1}^\infty\frac{(-1)^{sm+1}}{sm}e^{-sm w}e^{i(sm)(\alpha-\pi)/m}m\\
    &=\sum_{s=1}^\infty \frac{(-1)^{sm+1}}{s}e^{-rw}e^{i\alpha s}e^{-i \pi s},
\end{align}
recall that $m$ is even, hence $(-1)^{sm+1}e^{-i\pi s}=(-1)^{s+1}$. Now we can perform the summation over $s$ and obtain
\begin{equation}\label{eq:summation_p_result}
    \sum_{s=1}^\infty \frac{(-1)^{s+1}}{s}e^{-smw_k}e^{i\alpha s}=\ln\rbracket{1+e^{-mw_k}e^{i\alpha}}.
\end{equation}
Finally, combining Eq.~\eqref{eq:summation_p_trivial} with Eq.~\eqref{eq:summation_p_result} we have that 
\begin{equation}
    \sum_{p\in F_m}G(i\lambda_{p,\alpha})=\int_{}^{}\frac{\dd k}{2\pi}\sbracket{\ln\rbracket{1+e^{-mw_k}e^{i\alpha}}-m\ln\rbracket{1+e^{-w_k}}}.
\end{equation}
\section{Simplifying the integral expression for the $Z_m(\alpha)$} 
\label{equiv_results}
In this section, we show in details the simplifications occurring for the two-sites translational invariant state \eqref{eq:initial_state} in the asymptotic regime $t\gg x$. In particular we show that  the general expression in obtained in Sec.~\ref{equilibrium_case},  i.e. Eq.~\eqref{eq:summation_over_p}, simplifies to Eq.~\eqref{res_larget}.
In section \ref{subsecC1},  using symmetry arguments we show that 
\begin{align}\label{eq:simplified_expression}
\int_{-\pi}^{\pi}\frac{dk}{2\pi}H^{\alpha}_{m}(n_{k})&=\frac{i\alpha}{2}+\int_{-\pi}^\pi\frac{dk}{2\pi}\Re H^{\alpha}_{m}(n_{k}),\quad \text{with} \quad
H_{m}^{\alpha}(n_k)=\ln\left[ n_k^{m}e^{i\alpha}+(1-n_k)^{m}\right]. 
\end{align}
Note that for only this section the function $H_m^{\alpha}(k)$ defined in \eqref{eq:H_nk} is denoted as $H_m^{\alpha}(n_k)$ for convenience. In section~\ref{sectionC2}, we provide an alternative way to show Eq.~\eqref{eq:simplified_expression}, the proof is based on contour integration, in particular we show that
\begin{equation}
    \int_{-\pi}^\pi \frac{\dd k}{2\pi}\Im H_m^\alpha(n_k)=\frac{\alpha}{2},
\end{equation}
such that we recover Eq.~\eqref{eq:simplified_expression}.
\subsection{Symmetry based argument \label{subsecC1}}
For states of the form \eqref{eq:initial_state}, the occupation number is of the form Eq.~\eqref{occ_number_2sites}. 
\begin{align}
n_k&=\frac{1}{2}+\abs{a_0a_1^*}\cos\sbracket{k+\arg(a_0a_1^*)}. \label{eq:occupation_number}
\end{align}
As an intermediate step we show that one can rewrite the integral as 
\begin{align}
\int_{-\pi}^{\pi}\frac{dk}{2\pi}H_{m}^{\alpha}(n_k)=\int_{-\pi}^{\pi}\frac{dk}{2\pi}\frac{1}{2}\left[H_{m}^{\alpha}(1-n_k)+H_{m}^{\alpha}(n_k)\right]. \label{D4}
\end{align}
We note that the occupation number Eq.~\eqref{eq:occupation_number} verifies the property
\begin{align}
n_{k+\pi}&=\frac{1}{2}+\abs{a_0a_1^*}\cos\sbracket{k+\arg\rbracket{a_0a_1^*}+\pi}=\frac{1}{2}-\abs{a_0a_1^*}\cos\sbracket{k+\arg\rbracket{a_0a_1^*}}=1-n_k.\label{D8}
\end{align}
Hence, performing the change of variable $k\to k+\pi$, and using Eq.~\eqref{D8}, one can rewrite
\begin{align}
\int_{-\pi}^{\pi}\frac{dk}{2\pi}H_{m}^{\alpha}(n_k)&=\int_{0}^{2\pi}\frac{dk}{2\pi}H_{m}^{\alpha}(n_{k+\pi})=\int_{0}^{2\pi}\frac{dk}{2\pi}H_{m}^{\alpha}(1-n_k)\nonumber\\
&=\int_{-\pi}^{\pi}\frac{dk}{2\pi}H_{m}^{\alpha}(1-n_k),
\end{align}
where in the last line we used that $n_k$ is a $2\pi$ periodic function in $k$. This implies that integrating over $[-\pi,\pi]$ is the same as integrating over $[0,2\pi]$. Hence, we recover Eq.~\eqref{D4}. Notice,
\begin{align}
H_{m}^{\alpha}(1-n_k)&=\ln\left[(1-n_k)^{m}e^{i\alpha}+n_k^{m}\right]=\ln\left[(1-n_k)^{m}+n_k^{m}e^{-i\alpha}\right]+\ln(e^{i\alpha})\nonumber\\
&=\left[H_{m}^{\alpha}(n_k)\right]^{*}+i\alpha.
\end{align}
Hence, 
\begin{align}
\frac{H_{m}^{\alpha}(n_k)+H_{m}^{\alpha}(1-n_k)}{2}=\frac{H_{m}^{\alpha}(n_k)+\left[H_{m}^{\alpha}(n_k)\right]^{*}+i\alpha}{2}.
\end{align}
So that
\begin{align}
\Re\left[H_{m}^{\alpha}(n_k)\right]&=\frac{H_{m}^{\alpha}(n_k)+\left[H_{m}^{\alpha}(n_k)\right]^{*}}{2}=-\frac{i\alpha}{2}+\frac{H_{m}^{\alpha}(n_k)+H_{m}^{\alpha}(1-n_k)}{2}.\label{D13}
\end{align}
By integrating Eq.~\eqref{D13} and using Eq.~\eqref{D4}, we conclude that
\begin{align}
\int_{-\pi}^{\pi}\frac{dk}{2\pi}H_{m}^{\alpha}(n_k)=\frac{i\alpha}{2}+\int_{-\pi}^{\pi}\frac{dk}{2\pi}\Re\left[H_{m}^{\alpha}(n_k)\right].
\end{align}

\subsection{Contour integration argument \label{sectionC2}}
Alternatively, it is also possible to directly integrate the imaginary part and show that
\begin{align}
\int_{-\pi}^{\pi}\frac{dk}{2\pi}\Im H_{m}^\alpha\rbracket{n_k}=\frac{i\alpha}{2}.
\end{align}
First, substituting the imaginary part into the integral we have $H_m^\alpha(n_k)$
\begin{equation}
    \int_{-\pi}^{\pi} \frac{\dd k}{2\pi} \Im H_{m,\alpha}\rbracket{n_k} 
    =\int_{-\pi}^{\pi}  \frac{\dd k}{2\pi} \arctan \sbracket{
    \frac{n_k^m\sin\alpha}
    {n_k^m\cos\alpha+\rbracket{1-n_k}^m}}.
\end{equation}
We perform a change of variable $z=\exp\rbracket{ik+\arg{a_0a_1^*}}$ such that 
\begin{equation}
    \cos\sbracket{k+\arg\rbracket{a_0a_1^*}}=\frac{1}{2}\rbracket{z+\frac{1}{z}},
\end{equation}
then the integral in terms of $z$ is equivalent to a contour integral with radius 1 centered around the origin
\begin{equation}
    \int_{-\pi}^{\pi}  \frac{\dd k}{2\pi} \Im H_{m,\alpha}(k) =\oint  \frac{\dd z}{2\pi i} \frac{1}{z}\arctan \sbracket{\frac{\sbracket{\frac{1}{2}+\frac{\abs{a_0a_1^*}}{2}\rbracket{z+\frac{1}{z}}}^m\sin\alpha}{\sbracket{\frac{1}{2}+\frac{\abs{a_0a^*_1}}{2}\rbracket{z+\frac{1}{z}}}^m\cos\alpha+\sbracket{\frac{1}{2}-\frac{\abs{a_0a^*_1}}{2}\rbracket{z+\frac{1}{z}}}^m}}.
\end{equation}
The result of the integral is simply given by the residue of the integrand. We now introduce the following notation
\begin{equation}
    g_+ = \sbracket{\frac{1}{2}+\frac{\abs{a_0 a_1^*}}{2}\rbracket{z+\frac{1}{z}}}^m,\quad g_-= \sbracket{\frac{1}{2}-\frac{\abs{a_0 a_1^*}}{2}\rbracket{z+\frac{1}{z}}}^m.
\end{equation}
The argument of the $\arctan$ can be rewritten like
\begin{align}
    \frac{g_+\sin\alpha}{g_+\cos\alpha+g_-}&=\frac{g_+\sin\alpha}{g_+\cos\alpha} \rbracket{\frac{1}{1+\frac{g_-}{g_+\cos\alpha}}}\\
                                           &=\tan \alpha \sum_{n=0}^{\infty} (-\sec\alpha)^n \rbracket{\frac{g_-}{g_+}}^n.
\end{align}
One can now perform a series expansion for $g_-/g_+$ and obtain
\begin{equation}
    \frac{g_-}{g_+}=\sbracket{-1+O(z)}^m,
\end{equation}
recall that $m$ is even, hence $g_-/g_+=1+O(z)$, this is enough as we only need the coefficient of $z^0$ to extract the residue. Then 
\begin{align}
    \tan \alpha \sum_{n=0}^{\infty} (-\sec\alpha)^n \rbracket{\frac{g_-}{g_+}}^n&= \tan \alpha \sum_{n=0}^{\infty} (-\sec\alpha)^n \sbracket{1+O(z)}^n\\
&=\frac{\tan \alpha}{1+\sec\alpha}+O(z)\\
&=\tan \frac{\alpha}{2}+O(z)
\end{align}
then the original contour integral simplifies to 
\begin{align}
    \oint \frac{\dd z}{2\pi i} \frac{1}{z}\arctan\rbracket{\tan{\frac{\alpha}{2}+O(z)}}&=\oint \frac{\dd z}{2\pi i}\frac{\alpha}{2z}+O(1)\\
&=\frac{\alpha}{2},
\end{align}
where on the last line, the Cauchy-Residue theorem was used.

\section{Fourier transform of the initial state \label{Fourier_transform}}
Let us consider the two-sites translational invariant state Eq.~\eqref{eq_phi_c}
\begin{align}
|\Psi_0\rangle&=\prod_{j=1}^{L/2}\left[a_{0}\psi^{\dagger}(2j)+a_{1}\psi^{\dagger}(2j-1)\right]|0\rangle, & |a_{0}|^{2}+|a_{1}|^{2}=1, \label{eq_phi_c}
\end{align}
with the discrete Fourier transform 
\begin{align}
\psi^{\dagger}(j)=\frac{1}{\sqrt{L}}\sum_{k\in {\rm BZ}}e^{-ijk}\psi^{\dagger}(k).
\end{align}
Hence, we have
\begin{align}
|\Psi_0\rangle&=\prod_{j=1}^{L/2}\frac{1}{\sqrt{L}}\sum_{k\in {\rm BZ}}\left[a_{0}e^{-i2jk}\psi^{\dagger}(k)+a_{1}e^{-i(2j-1)k}\psi^{\dagger}(k)\right]|0\rangle,\nonumber\\
&=\prod_{j=1}^{L/2}\frac{1}{\sqrt{L}}\sum_{k\in {\rm BZ}}e^{-i2jk}\left[a_{0}+a_{1}e^{ik}\right]\psi^{\dagger}(k)|0\rangle,\nonumber\\
\end{align}
In the end
\begin{align}
|\Psi_0\rangle&=\prod_{j=1}^{L/2}\frac{1}{\sqrt{L}}\left\{\sum_{k\in {\rm RBZ}}e^{-i2jk}\left(a_{0}+a_{1}e^{ik}\right)\psi^{\dagger}(k)+\sum_{k\in {\rm RBZ}}e^{-i2jk}\left(a_{0}+a_{1}e^{ik+i\pi}\right)\psi^{\dagger}(k+\pi)\right\}|0\rangle,\nonumber\\
&=\prod_{j=1}^{L/2}\frac{1}{\sqrt{L}}\left\{\sum_{k\in {\rm RBZ}}e^{-i2jk}\left[\left(a_{0}+a_{1}e^{ik}\right)\psi^{\dagger}(k)+\left(a_{0}+a_{1}e^{ik+i\pi}\right)\psi^{\dagger}(k+\pi)\right]\right\}|0\rangle.
\end{align}
Let us now denote
\begin{align}
\chi^{\dagger}_{k}=\left[\left(a_{0}+a_{1}e^{ik}\right)\psi^{\dagger}(k)+\left(a_{0}+a_{1}e^{ik+i\pi}\right)\psi^{\dagger}(k+\pi)\right].
\end{align}
One can expand the product. However, given that we consider fermionic operators, one has $\chi^{\dagger}_{k}\chi^{\dagger}_{k'}=-\chi^{\dagger}_{k'}\chi^{\dagger}_{k}$. Hence, if $k_{i}=k_{j}$ for $i\neq j$, then the contribution vanishes. One can then write the product as a sum over the permutations of distinct momenta $k_{i}\neq k_{j}$, $\forall i,j$
\begin{align}
\prod_{j=1}^{L/2}\frac{1}{\sqrt{L}}\left\{\sum_{k\in {\rm RBZ}}e^{-i2jk}\chi^{\dagger}_{k}\right\}|0\rangle=\left(\frac{1}{\sqrt{L}}\right)^{L/2}\sum_{k_{1}\neq k_{2}\neq\cdots\neq k_{N}\in {\rm RBZ}}\left[\prod_{j=1}^{L/2}e^{-i2jk_{j}}\right]\chi^{\dagger}_{k_{1}}\chi^{\dagger}_{k_{2}}\cdots\chi^{\dagger}_{k_{L/2}}|0\rangle
\end{align}
We now re-order the sum into fixed sector $k_{1}<k_{2}<\cdots<k_{L/2}$. This corresponds to permute the fermionic operators that induces a sign of the permutation
\begin{align}
\prod_{j=1}^{L/2}\frac{1}{\sqrt{L}}\left\{\sum_{k\in {\rm RBZ}}e^{-i2jk}\chi^{\dagger}_{k}\right\}|0\rangle\nonumber&=\left(\frac{1}{\sqrt{L}}\right)^{L/2}\sum_{k_{1}<k_{2}<\cdots<k_{L/2} \in {\rm RBZ}}\sum_{\sigma\in S_{L/2}}{\rm sgn}(\sigma)\left[\prod_{j=1}^{L/2}e^{-i2jk_{\sigma(j)}}\right]\nonumber\\
&\times\chi^{\dagger}_{k_{1}}\chi^{\dagger}_{k_{2}}\cdots\chi^{\dagger}_{k_{L/2}}|0\rangle,
\end{align}
where $S_{L/2}$ is the ensemble of permutations of the momenta of the RBZ $(k_{1},k_{2},\cdots,k_{L/2})\in {\rm RBZ}$. Furthermore because there are only $L/2$ terms in the RBZ, the summation $\sum_{k_{1}<k_{2}<\cdots<k_{L/2} \in {\rm RBZ}}$ only contains one term. Setting 
\begin{align}
M_{j\alpha} = e^{-i 2 j k_a}, \quad j,a = 1,\dots,L/2,
\end{align}
one can use the Leibniz formula to express the previous expression as a determinant
\begin{align}
\sum_{\sigma \in S_{L/2}} \mathrm{sign}(\sigma) \prod_{j=1}^{L/2} e^{-i 2 j k_{\sigma(j)}}=\sum_{\sigma \in S_{L/2}} \mathrm{sign}(\sigma) \prod_{j=1}^{L/2} M_{j,\sigma(j)}=\det(M).
\end{align}
Hence, the wavefunction reduces to a Slater determinant that takes into account the symmetrization properties of the fermionic wavefunction recovering Eq.~\eqref{Fourier_transform_2_sites}.
\section{Poisson-Binomial Process and Full Counting Statistics} \label{Poisson-Binomial}

In this section we provide a microscopic derivation of the charged moments $Z_m(\alpha)$ and show how its relation to the Poisson-Binomial process. In particular we consider a density matrix $\rho|_A$ at finite volume $\abs{A}$ and show that it can characterize the distribution of $N_A$ (the particle number operator restricted to $A$) in the same way as $\rho_A$, where $\rho_A$ is the true RDM associated with a GGE. To define the density matrix at finite volume $\abs{A}$ we introduce the fermionic fields at finite volume $\Psi(x)$ which are still in the continuum.
We consider boundary conditions $\Psi(x+\abs{A})=\Psi(x)$, as such the field $\Psi(x)$ allows a Fourier series decomposition.
\begin{equation}
    \Psi(x)=\sum_{k}e^{ikx}\Psi_k,\quad k\in \frac{2\pi}{L}\Z,
\end{equation}
where the $\Psi_k$ are the Fourier modes
\begin{equation}
    \Psi_k = \frac{1}{L}\int_{0}^{L} e^{-ikx}\Psi(x) \dd x,
\end{equation}
The charge $N_A$ in this basis is 
\begin{equation}
    N_A=\int_{0}^{L} \Psi^\dagger(x)\Psi(x) \dd x=\sum_{k} N_k, \quad N_k = \Psi^\dagger_k\Psi_k.
\end{equation}
We now treat the eigenvalues of $N_k$ as a Bernoulli random variable, whose outcomes' probabilities are given in terms of its eigenvalues
\begin{equation}
    P(N_k = 1) = n_k,\quad P(N_k=0)=1-n_k.
\end{equation}
%
The probability distribution of $N_A$  can be recovered by making use of the following density matrix for the system at finite volume $\abs{A}$ 
\begin{equation}
    \rho|_A = \bigotimes_k \rho|_A(k), \quad  \rho|_A(k) = \rbracket{1-n_k}\ket{0}\bra{0}+ n_k \ket{1}\bra{1}.
\end{equation}
Furthermore we denote the normalized $m$ copy density matrix as $\rho^{(m)}|_A$ 
\begin{equation}
    \rho^{(m)}|_A \equiv \frac{(\rho|_A)^m}{\tr (\rho|_A)^m} = \bigotimes_k \sbracket{\rbracket{1-n^{(m)}_k}\ket{0}\bra{0}+ n_k^{(m)}\ket{1}\bra{1}}
\end{equation}
which is of the same form of $\rho|_A$, but with an $m$-dependent occupation function
\begin{equation}
    n_k^{(m)} = \frac{n_k^m}{n_k^m+(1-n_k)^m}.
\end{equation}
Now computing the FCS of $N_A$ using $\rho^{(m)}|_A$ one obtains
\begin{align}
    \tr \sbracket{\rho^{(m)}|_A e^{i\alpha N_A}} &= \prod_k\sbracket{1-n^{(m)}_k\tr\rbracket{\ket{0}\bra{0}e^{i\alpha N_k}}+n^{(m)}_k \tr\rbracket{\ket{1}\bra{1}e^{i\alpha N_k}}}\\
                                                 &=\prod_k\sbracket{ n_k^{(m)}e^{i\alpha}+1-n_k^{(m)}}.
\end{align}
We can then take the thermodynamic limit and recover $Z_m(\alpha)/Z_m(0)$. Hence, we have demonstrated that the fluctuations of $N_A$ are characterized by the RDM can equivalently be characterized by the density matrix $\rho^{(m)}|_A$ and that the FCS is equivalent to the one of a Poisson-Binomial process. 
This mapping allows to get exact results for the Fourier transform as well, and the result is known and given by the probability distribution 
\begin{equation}
    P^{(m)}(N_A = q)=
    \sum_{k_1<k_2<\ldots<k_{n_a}}\prod_{j=1}^{q} n^{(m)}_{k_j}\prod_{p\neq k_1\neq k_2\neq\ldots\neq k_{q}}\rbracket{1-n^{(m)}_{p}}.
\end{equation}
Notice that in the thermodynamic limit this would give $S_m (q)$, up to a normalization constant.

\end{appendix}

\bibliography{references.bib}
\end{document}